\documentclass[aps,prl,reprint, superscriptaddress]{revtex4-2} 
\usepackage{amsmath,amssymb,amsfonts}
\usepackage{graphicx}
\usepackage[dvipsnames]{xcolor}
\usepackage{comment}
\usepackage[overload]{empheq}
\usepackage{upgreek}
\usepackage{lmodern,pdftexcmds}
\usepackage[sort&compress]{natbib}
\usepackage{hyperref}
\usepackage{soul}
\usepackage[normalem]{ulem}
\usepackage{enumerate}

\usepackage[mathlines]{lineno}
\usepackage{etoolbox} 

\newcommand*\linenomathpatch[1]{%
	\cspreto{#1}{\linenomath}%
	\cspreto{#1*}{\linenomath}%
	\csappto{end#1}{\endlinenomath}%
	\csappto{end#1*}{\endlinenomath}%
}

\expandafter\ifx\linenomath\linenomathWithnumbers
\let\linenomathAMS\linenomathWithnumbers
\patchcmd\linenomathAMS{\advance\postdisplaypenalty\linenopenalty}{}{}{}
\else
\let\linenomathAMS\linenomathNonumbers
\fi

\linenomathpatch{equation}

\usepackage{float}
\usepackage{amsmath,amsfonts,amsthm} 
\usepackage{bm}

\usepackage{appendix}
\usepackage{subeqnarray}
\usepackage{color}
\usepackage{mathtools}

\newcommand{\bs}{\boldsymbol}
\newcommand{\dbs}[1]{\dot{\boldsymbol{#1}}}
\newcommand{\tn}{\wideoverline{\bs{\nabla}}}
\newcommand{\on}{\bs{\nabla}}
\newcommand{\wb}[1]{\wideoverline{\bs{{#1}}}}
\newcommand{\bbb}[1]{\overline{\bs{#1}}}
\newcommand{\obb}[1]{\hat{\overline{#1}}}
\newcommand{\oa}[1]{\wideoverline{\mathcal{#1}}}
\newcommand{\rb}[1]{\bs{\mathcal{#1}}}
\newcommand{\pf}[2]{\frac{\partial #1}{\partial #2}}
\newcommand{\pfw}[2]{\frac{\partial \wideoverline{#1}}{\partial \wideoverline{#2}}}
\newcommand{\pfm}[3]{\frac{\partial^#1 #2}{\partial #3^#1}}
\newcommand{\pft}[2]{\frac{\partial^2 #1}{\partial #2^2}}
\newcommand{\pfh}[2]{\frac{\partial^3 #1}{\partial #2^3}}
\newcommand{\df}[2]{\frac{\mathrm{d} #1}{\mathrm{d}  #2}}
\newcommand{\dfm}[3]{\frac{\mathrm{d} ^#1 #2}{\mathrm{d}  #3^#1}}
\newcommand{\dft}[2]{\frac{\mathrm{d} ^2 #1}{\mathrm{d}  #2^2}}
\newcommand{\md}{\mathrm{d}}
\newcommand{\f}{p^\text{off}}
\newcommand{\n}{p^\text{on}}

\newcommand{\fw}{P^\text{off}}
\newcommand{\nw}{P^\text{on}}

\newcommand{\m}{p^\text{m}}
\newcommand{\pd}{p^\text{d}}
\newcommand{\mw}{P^\text{m}}
\newcommand{\dw}{P^\text{d}}
\newcommand{\ab}[1]{\langle #1 \rangle}
\newcommand{\abp}[1]{\langle #1 \rangle_{\bs{p}}}
\newcommand{\abpp}[1]{\langle #1 \rangle_{\bs{pp}}}
\newcommand{\re}{\rho_\text{eq}}
\newcommand{\R}{R_\text{eq}}
\newcommand{\w}[1]{\wideoverline{#1}}
\newcommand{\pe}{\mathrm{Pe}}
\newcommand{\nb}{\bs{\nabla}_{\bs{x}}}
\newcommand{\np}{\bs{\nabla}_{\bs{p}}}
\newcommand{\Dt}[1]{\frac{\mathrm{D} #1}{\mathrm{D} t}}

\newcommand{\ob}[1]{\overline{\boldsymbol{{#1}}}}
\newcommand{\obn}{\ob{\nabla}}
\renewcommand{\ol}[1]{\overline{#1}}

\begin{document}

\title{ Geometry-Dependent Defect Merging Induces Bifurcated Dynamics in Active Networks}
\author{Fan Yang}
\email{fy2@caltech.edu}
\affiliation{Division of Biology and Biological Engineering, California Institute of Technology, Pasadena, CA, USA}

\author{Shichen Liu}
\affiliation{Division of Biology and Biological Engineering, California Institute of Technology, Pasadena, CA, USA}

\author{Hao Wang}
\affiliation{Division of Biology and Biological Engineering, California Institute of Technology, Pasadena, CA, USA}

\author{Heun Jin Lee}
\affiliation{Department of Applied Physics, California Institute of Technology, Pasadena, CA, USA}

\author{Rob Phillips}

\affiliation{Division of Biology and Biological Engineering, California Institute of Technology, Pasadena, CA, USA}

\affiliation{Department of Applied Physics, California Institute of Technology, Pasadena, CA, USA}

\author{Matt Thomson}
\email{mthomson@caltech.edu}

\affiliation{Division of Biology and Biological Engineering, California Institute of Technology, Pasadena, CA, USA}

\pacs{}

\begin{abstract}
	Cytoskeletal networks can repair defects to maintain structural integrity. However, the mechanisms and dynamics of defect merging remain poorly understood. Here we report a geometry-tunable merging mechanism in microtubule-motor networks initiated by active crosslinking. We directly generate   defects using a light-controlled microtubule-motor system in O-shaped and V-shaped networks, and observe that the defects can self-close. Combining theory and experiment, we find that the  V-shaped networks must overcome  internal elastic resistance in order to zip up cracks, giving rise to a bifurcation of dynamics dependent on the initial opening angle of the crack: the crack  merges below a critical angle and opens up at larger angles. Simulation of a continuum model reproduces the bifurcation dynamics, revealing the importance of overlapping boundary layers where free motors and microtubules can actively crosslink and thereby merge the defects.   We  also formulate a simple elastic-rod model that can qualitatively predict the critical angle, which is  tunable by the network geometry.

\end{abstract}

\maketitle

Cytoskeletal networks can dynamically reconfigure themselves and generate force to fulfill crucial functions in life, such as mechanical support, motility, and division of cells. Furthermore, in neurons, microtubules are organized into parallel arrays that serve as tracks for cargo transport \cite{baas11}. After an axon is injured, the rearrangement of microtubule orientations into parallel arrays plays a key role in axon regeneration \cite{he16}. Kinesin motors that can actively bind and walk on microtubules may contribute to the healing and alignment of microtubule networks \cite{braun09}. However, such mechanisms of healing by motors are not well studied. Previous research on self-healing cytoskeletal networks has mainly focused on mechanisms through adding or reassembling the subunits that make up the cytoskeleton. For example, individual microtubules are found to be  capable of incorporating free tubulins to repair lattice defects \cite{schaedel15}. At the network level, filamentous actin hydrogels can restore their storage modulus through dynamic polymerization and depolymerization of globular actin, after a shear strain is removed \cite{sano11}.  Motor proteins can also reconnect laser-ablated microtubule bundles in mitotic spindles \cite{elting14}. In this Letter, we investigate how active crosslinking by motor proteins drives geometry-dependent defect merging in O-shaped and V-shaped microtubule networks, leading to bifurcated dynamics.

\begin{figure*}[t]
	\includegraphics[width=\linewidth,angle=0]{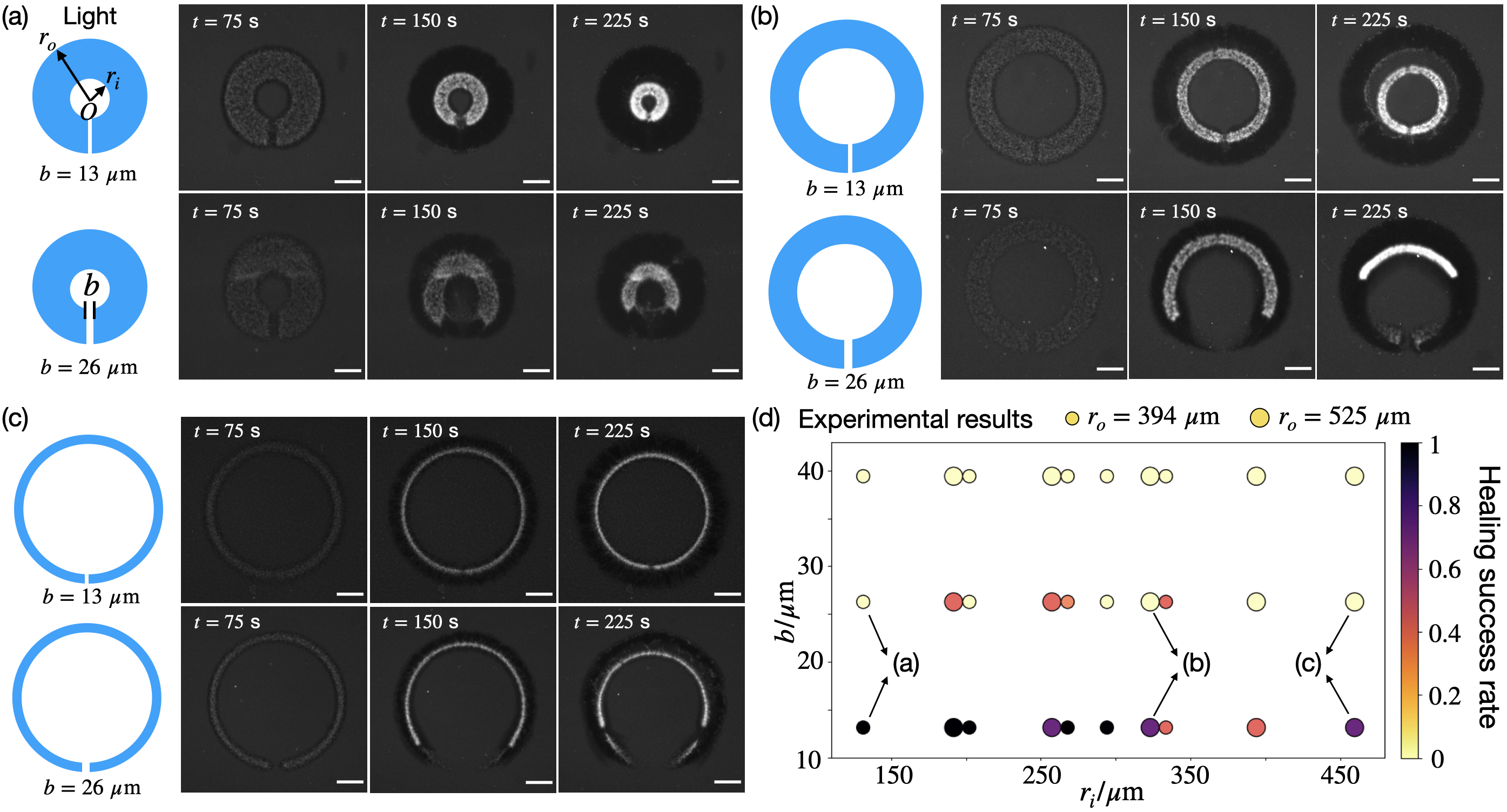}
	\caption{Self-healing behavior of O-shaped active networks is governed by a critical gap width. (a-c) Experimental images of O-shaped networks with gap defects. In each panel, the inner and outer radii are fixed, while two initial gap widths, $b=13$ $\mu$m and 26 $\mu$m, are shown. Scale bars, 200 $\mu$m. (d) Experimental measurements of healing success rates with varying gap widths $b$ and inner radii $r_i$.  Small and large markers correspond to outer radii $r_o = 394$ $\mu$m and 525 $\mu$m, respectively. Experiments (a-c) are labeled on the diagram.} 
	\label{o-ring}
\end{figure*}


Our reconstituted microtubule-motor system \cite{ross19, yang24} provides  a light-controllable platform with minimum components that can generate self-healing networks and elucidate the underlying mechanisms. The experimental  system \cite{yang24} consists of free microtubules, light-activatable motor proteins, ATP and buffer solutions, placed in a flow cell, whose height, around 100 $\mu$m, is much smaller than its horizontal dimensions (Supplemental Material).   Depolymerization and polymerization of tubulins can be neglected in our experiments. The microtubules are stabilized  to minimize  depolymerization \cite{ross19}. Polymerization doubles the average microtubule length, initially around 1.3 $\mu m$, every 4 hours, which  is very slow compared to the healing dynamics at the scale of minutes. The engineered motor proteins can  ``link" under blue light. We use the terms ``linked" and ``unlinked" motors to distinguish these two states. Microtubule networks of arbitrary shapes can be generated through  light projections onto the flow cell. The networks are contractile due to crosslinking by motors. 

We directly generate O-shaped networks with defects to investigate whether they can self-heal.  As shown in Fig. \ref{o-ring}(a-c), when the gap width $b$ is small, the defect  merges and the O-shaped network contracts as a whole. In contrast, when $b$ is large, the defect expands, leading to the opening of the O-shaped network. The dynamics—either opening or closing of the gap—is decoupled from the overall contraction of the network. The healing success rates, defined as the percentage of successful merging experiments among total experimental replicates, are documented in Fig. \ref{o-ring}(d) with varying geometrical parameters. Our experiments reveal a consistent critical gap threshold $b_c$, within the range of 13--26 $\mu$m, that governs the self-healing behavior of O-shaped networks. Across various inner and outer radii, the O-shaped network tends to merge successfully when $b<b_c$ and fail to merge when $b>b_c$.

We measure the light intensity across the gap and find that the critical gap threshold, $b_c$, is determined by the effective activation region of the projected light. At the edges of the projected light pattern, the light intensity decays to the background level within a $\sim$ 20 $\mu$m layer, as shown in Fig. S7 of the Supplemental Material. When the projected gap width is $b = 13$ $\mu$m, the two opposing light decay regions strongly overlap, whereas for $b = 26$ $\mu$m the overlapping region is small. We hypothesize that these light decay regions give rise to a boundary layer of linked motors adjacent to the defect interfaces. The gap  closes only when the boundary layers from opposite sides significantly overlap, allowing the linked motors within the overlapping region to crosslink microtubules and merge the gap.

\begin{figure*}[t]
	\includegraphics{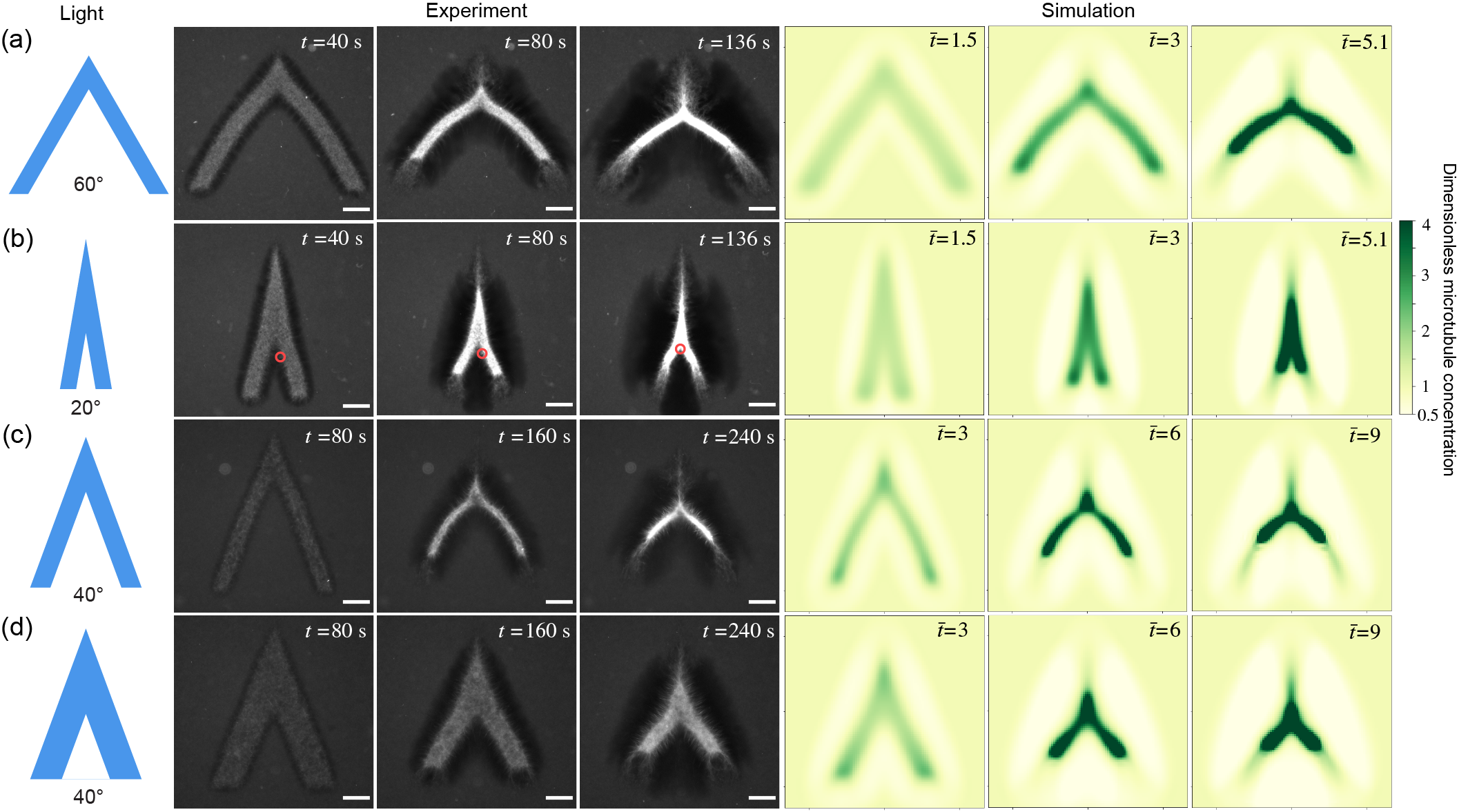}
	\caption{Experiments and simulations of  V-shaped active networks show a bifurcation of merging- and buckling-dominated dynamics dependent on the network geometry. (a) and (b) are two networks with the same initial arm lengths and widths but different opening angles. The networks buckle at  the large opening angle (a) while merge at the small angle  (b). Red circles in (b) track a small protrusion on the right arm which eventually merges with the left arm, demonstrating the partial closure of the crack. (c) and (d) are two networks of the same arm lengths and opening angles but different arm widths. The thinner network (c) buckles outwards while the thicker one (d) bends inwards. The spatiotemporal dimensions of the simulated and  experimental images are matched. Simulation details are in the Supplemental Material. $t$ ($\overline{t}$) is (dimensionless) time after the first light pulse. In simulations, the microtubule concentration is non-dimensionalized by the initial microtubule concentration. Scale bar, 100 $\mu$m. } 
	\label{exp}
\end{figure*}

To further test the boundary-layer hypothesis, we create V-shaped networks to mimic cracks, and find that there exists a critical initial opening angle above which the network buckles, and below which it merges.  Fig. \ref{exp}(a) and \ref{exp}(b) show two networks with the same initial arm lengths and widths but different opening angles. The network with the larger angle in Fig. \ref{exp}(a) keeps opening up as it contracts. Its two arms bend outwards and form a convex shape. In contrast, the network  with the smaller initial angle in Fig. \ref{exp}(b) closes in and the two arms zip up, forming a concave shape.  The critical opening angle also depends on the network geometry. We generate two networks with fixed arm lengths and opening angles but different widths, as shown in Fig. \ref{exp}(c) and \ref{exp}(d), and find that the thinner network buckles outwards while the thicker one bends inwards, indicating that the critical angle can be tuned by the arm shape.

\begin{figure}[h]
	\includegraphics[width=1\linewidth,angle=0]{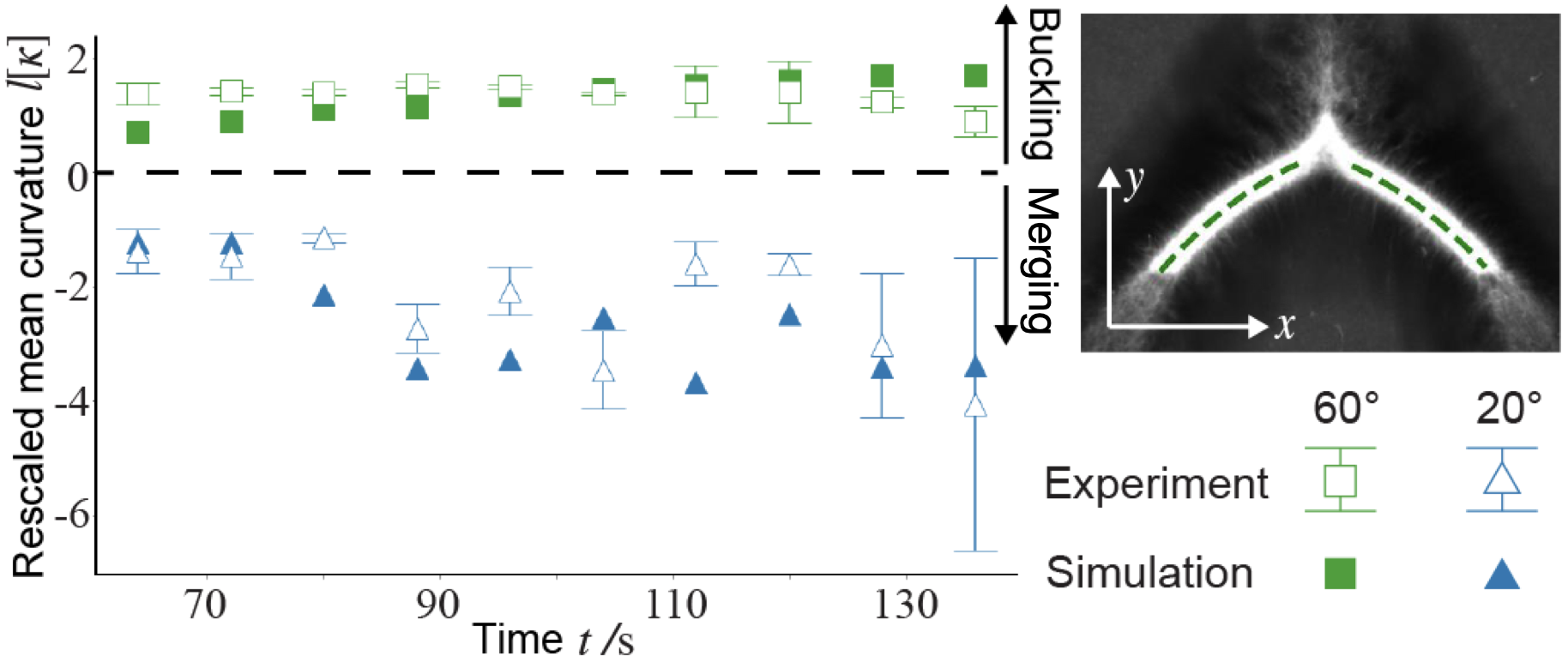}
	\caption{Merging and buckling-dominated dynamics can be differentiated by the network curvature. The curvature is negative (concave) for the former and positive (convex) for the latter. The mean curvature $[\kappa]$ is averaged over the centerline of each arm excluding the tip region (dotted lines in the upper right inset), and rescaled by the initial arm length $l$. Error bars represent the difference between left and right arms in a single experiment.  }
	\label{curv}
\end{figure}

The two distinct phenomena in Fig. \ref{exp}(a) and \ref{exp}(b) demonstrate a bifurcation of the active network dynamics dependent on the initial opening angles. We denote the dynamics in Fig. \ref{exp}(a) and Fig. \ref{exp}(b) as the buckling-dominated and  merging-dominated regimes, respectively. The two regimes can be quantitatively distinguished  by curvature of the network. Given a centerline profile $y(x)$ (inset in Fig. \ref{curv}), the local curvature $\kappa$ is defined as $\kappa =y''/(1+y'^2)^{3/2}$.   We define the mean  curvature $[\kappa]$ along the centerline as $[\kappa]= \int_{x_0}^{x_t} \frac{y''}{1+y'^2} \mathrm{d} x / \int_{x_0}^{x_t}\sqrt{1+y'^2} \mathrm{d} x$, where $x_0$ and $x_t$ denote the starting and ending points of the centerline, respectively.  Time evolutions of mean curvatures   in Fig. \ref{exp}(a) and \ref{exp}(b) are plotted in Fig. \ref{curv}. By our definition,  negative curvature represents a concave shape--characteristic of the merging-dominated regime. Additional images of concave networks are provided in Fig. S8 of the Supplemental Material.
Conversely, positive curvature corresponds to a convex shape, typical of the buckling-dominated regime, where the two arms bend outward.


We conduct numerical simulations to uncover  the zipping-up mechanisms.  The simulations in Fig. \ref{exp} are based on a three-phase model from our previous work \cite{yang24} (Supplemental Material). The simulation can reproduce the bifurcation dynamics (Fig. \ref{exp}) and the simulated curvatures are in qualitative agreement with experiments (Fig. \ref{curv}]).  We also compute a bifurcation phase diagram by varying crosslinking rates and find that increasing the crosslinking rate promotes the network merging, thereby confirming that the merging process is driven by active crosslinking (Supplemental Material).

\begin{figure}[h]
	\includegraphics[width=0.9\linewidth,angle=0]{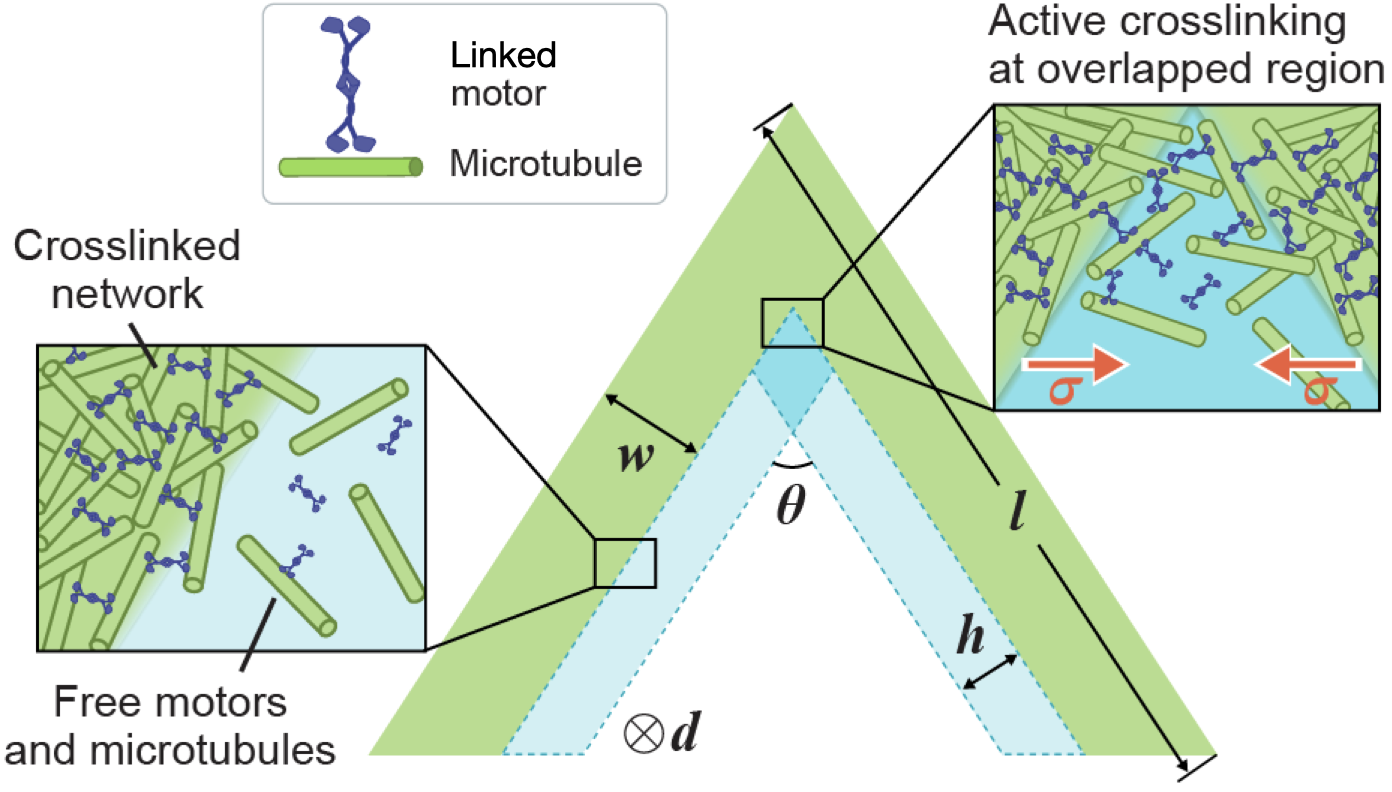}
	\caption{Schematic of the elastic-rod model. The crosslinked network is modeled as a kinked elastic rod (green). There are  boundary layers (light blue) of free motors and microtubules next to the network surfaces. Active crosslinking takes place at the overlapped region (dark blue) of the two boundary layers. }
	\label{ermodel}
\end{figure}

The boundary-layer hypothesis can  explain the opening-angle dependency of the bifurcation. As $\theta$ decreases, the overlapped region expands, favoring merging at small angles.  In this merging-dominated regime, the overlapped region becomes a ``zipping front" that can propagate along and zip up the two network arms.  It is well known that an elastic rod buckles when the compressive load exceeds a critical threshold. Buckling has also been reported in rectangular active networks \cite{ross19}. The V-shaped network can be viewed as a joint of two rectangular segments, where such buckling may occur. We propose that the opening dynamics observed in Fig. \ref{exp}(a) and \ref{exp}(c) results from a buckling instability driven by the compressive active stress. Once buckling initiates, the two arms will only bend outwards  to reduce the bending energy, which scales quadratically with local curvature \cite{audoly10}, at the kink.

Based on the boundary-layer and buckling hypotheses, we propose a simple elastic-rod model to predict the critical angle in bifurcation. We treat the crosslinked network as an elastic rod, as shown in Fig. \ref{ermodel}. The active stress, denoted by $\sigma_a$, generates a compressive force in each arm that is $F_a = \sigma_a w d$, with $w$ and $d$  the arm width and depth, respectively.   Within the overlapped boundary layer, active crosslinking induces an attractive force $F_h$ between the two arms. Assuming a surface force density $\sigma$ (inset in Fig. \ref{ermodel}),   the  attractive force on each arm is given by $F_h=\sigma d h / \sin \theta$, where  $h$ is the  boundary layer width. The tangential component of $F_h$ along each arm  introduces a tension  and the minimum compression within each arm is $F_c \approx  \sigma_a w d -\frac{\sigma d h}{2 \sin (\theta/2)} $. The bifurcation behavior of the network arises from competition between two mechanisms:  buckling, characterized by $F_c$, and merging, characterized by $F_h$.  We   estimate the critical buckling load using the classical Euler's result,  $F_b=\pi^2 EI/l^2$ \cite{howell}, where $E$ is the Young's modulus, and $I = d w^3/12$ is the moment of inertia.  The critical angle $\theta^*$ is determined by $F_c=F_b$, which is
\begin{equation}
	\theta^* = 2\arcsin \frac{\sigma h}{2w\left(\sigma_a-C  E \delta^2\right)}, \label{theta}
\end{equation}
where $\delta = w/l$ is the aspect ratio of each arm and $C= \pi^2/12$ is a constant. The network will merge when $\theta < \theta^*$ and buckle when  $\theta > \theta^*$.  From  (\ref{theta}), it follows that $\theta^*$ can be tuned by two dimensionless geometric parameters: the ratio of  boundary layer width to arm width $h/w$, and  the aspect ratio of the arm $\delta= w/l$. Both $w$ and $\delta$ are programmable through light signals in experiments and simulations. When $\delta$ is fixed, increasing $w$ will decrease the critical angle $\theta^*$. This is because the healing force $F_h$ depends only on $\theta$ and is independent of $w$ and $\delta$, whereas the compression $F_c$ in each arm scales linearly with $w$. As the arm width increases, a smaller $\theta^*$ is required to produce a greater overlapped region, and consequently a larger $F_h$, to balance the increasing compression. Conversely, fixing $w$ and decreasing the aspect ratio  $\delta$  also reduces $\theta^*$. In this case,  $F_c$ remains unaffected by  $\delta$, but a smaller $\delta$ will make the network more slender and prone to buckling. Therefore, a smaller $\theta^*$ is needed to generate sufficient attractive force $F_h$ to suppress the elastic instability.  Finally, we note that there is always overlapping of boundary layers in the tip region, rendering it always concave. The convexity and concavity predicted by our simple model  (\ref{theta}) apply only to the bulk region away from the tip.

\begin{figure}
	\includegraphics[width=1\linewidth,angle=0]{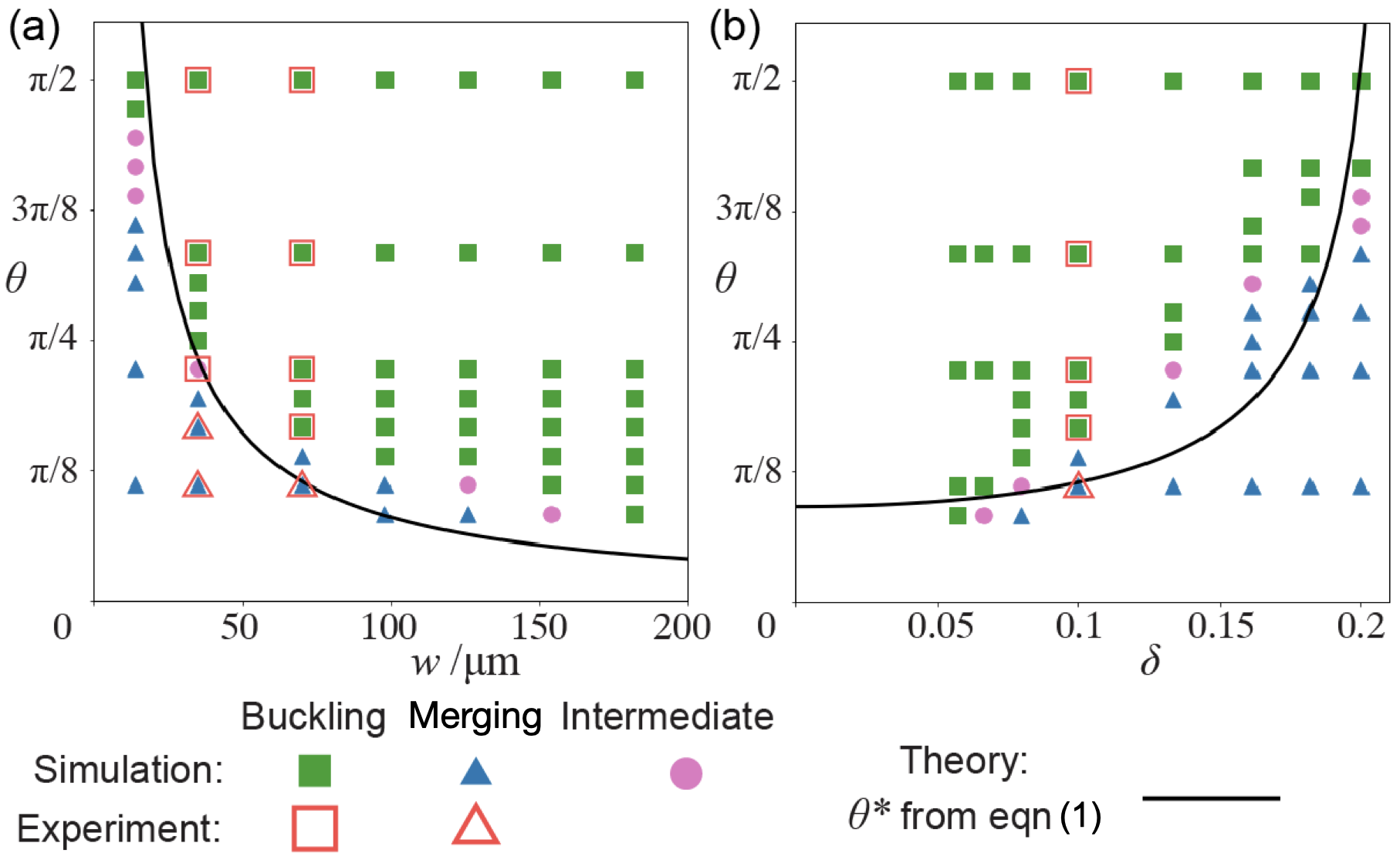}
	\caption{Bifurcation phase diagrams can be qualitatively predicted by the elastic-rod model. We fix $\delta = 0.1$ in (a) and $w=70$ $\mu$m in (b).  The fitting parameters used to plot the theoretical curves are $\sigma h/\sigma_a = 20$ $\mu$m and $\sigma h/C E= 1$ $\mu$m. ``Intermediate" represents when the network does not buckle and also does not show significant merging, such as Fig. \ref{exp}(d). In simulations,  the ``Intermediate"  state is characterized by the average curvature close to 0.}
	\label{pd}
\end{figure}

To test how network geometry can tune the bifurcation dynamics, we perform simulations with varying opening angles $\theta$, arm width $w$, and aspect ratio $\delta$. The outcomes---whether the network buckles or merges---are documented in the bifurcation diagrams in Fig. \ref{pd}.   We first fix $\delta=0.1$ vary the arm width  $w$ in Fig. \ref{pd}(a). As $w$ increases, the critical angle $\theta^*$ decreases. This confirms our theory (\ref{theta}) that increasing network size while preserving shape makes buckling easier and merging more difficult. This is because the healing force $F_h$ does not scale with the network size, whereas the active compression $F_c$ increases linearly with $w$. Consequently, as the network size grows, characterized by increasing $w$ when $\delta$ is fixed, the merging effect becomes less dominant. Similarly, Fig. \ref{pd}(b) shows the bifurcation phase diagram for varying $\delta$ at fixed $w$, which is again consistent with our theory:  as the aspect ratio $\delta$ increases, the network arms become shorter and more resistant to buckling. Both simulated phase diagrams are in quantitative agreement with experimental results. In general, $h$, $\sigma$, $\sigma_a$ and $E$ in our theory (\ref{theta}) depend on local microtubule and motor concentrations. We treat them as constants in plotting the theory in Fig. \ref{pd} for simplicity. Even so, the elastic-rod model can qualitatively predict the  bifurcation phase diagram and offers a clear explanation for its dependence on $w$ and $\delta$.  Furthermore, we can define two dimensionless groups, $A= 2 w \sigma_a / \sigma h$ and $B= 2CE\delta^2 w/\sigma h$,  and rewrite  (\ref{theta})  as $\theta^* = 2 \arcsin \left[(A- B)^{-1}\right]$, where $A$ is the ratio of  active compression to merging force, and $B$ is the ratio of  critical buckling load to  merging force. Increasing $A$ or decreasing $B$ makes the network easier to buckle, thereby reducing $\theta^*$.  As an anonymous reviewer pointed out, the quantitative discrepancy between our simple model and the linear dependence of $\theta^*$ on $\delta$ predicted by simulations in Fig. \ref{pd}b may also arise from the omission of effective surface tension in (1). This surface tension, arising from the motor activity, may cause the active network to bend inwards to reduce surface area \cite{yang20}--conceptually analogous to the barreling instability described in Ref. \cite{oriola20}.

In summary,  we  show that  overlapping of motor boundary layers can merge defects in active networks. For V-shaped cracks, the active crosslinking also needs to overcome an elastic instability which will open up the crack, leading to a bifurcation of merging and buckling that can be tuned by the initial network geometry. It has been increasingly evident that cytoskeletal networks are gel-like materials \cite{ahmed18} and vulnerable to a plethora of mechanical instabilities driven by self-generated active forces. Another example is the bifurcation of in-plane bending and out-of-plane buckling instabilities found in extensile active  sheets \cite{najma22}. However, instabilities are not always detrimental. Cells can  regulate and exploit mechanical instabilities to form functional structures, such as  mitotic spindles which are  shaped by a barreling-type instability \cite{oriola20}. Further work is needed to complete a mechanical-instability phase diagram of active networks and to uncover the  regulatory mechanisms used by cells to control such instabilities.



\begin{acknowledgements}
	We are grateful to Zhen-Gang Wang and Howard A. Stone for fruitful discussions and to Inna Strazhnik for illustration. This project is funded by NIH 1R35 GM118043 (MIRA), Packard Foundation, Moore Foundation, and Donna and Benjamin M. Rosen Bioengineering Center. FY also acknowledges  support from BBE Divisional Postdoctoral Fellowship at Caltech.
\end{acknowledgements}

\bibliography{DefectMerging, selforg, FlowControlref}

\newpage
\onecolumngrid

\begin{center}
\textbf{\large Supplemental Material}
\end{center}


\setcounter{equation}{0}
\setcounter{figure}{0}
\setcounter{table}{0}
\setcounter{section}{0}
\setcounter{page}{1}

\makeatletter
\renewcommand{\theequation}{S\arabic{equation}}
\renewcommand{\thefigure}{S\arabic{figure}}

\renewcommand{\bs}{\boldsymbol}
\renewcommand{\dbs}[1]{\dot{\boldsymbol{#1}}}
\renewcommand{\tn}{\widetilde{\bs{\nabla}}}
\renewcommand{\on}{\bs{\nabla}}
\renewcommand{\wb}[1]{\widetilde{\bs{{#1}}}}
\renewcommand{\bbb}[1]{\overline{\bs{#1}}}
\renewcommand{\obb}[1]{\hat{\overline{#1}}}
\renewcommand{\oa}[1]{\widetilde{\mathcal{#1}}}
\renewcommand{\rb}[1]{\bs{\mathcal{#1}}}
\renewcommand{\pf}[2]{\frac{\partial #1}{\partial #2}}
\renewcommand{\pfw}[2]{\frac{\partial \widetilde{#1}}{\partial \widetilde{#2}}}
\renewcommand{\pfm}[3]{\frac{\partial^#1 #2}{\partial #3^#1}}
\renewcommand{\pft}[2]{\frac{\partial^2 #1}{\partial #2^2}}
\renewcommand{\pfh}[2]{\frac{\partial^3 #1}{\partial #2^3}}
\renewcommand{\df}[2]{\frac{\mathrm{d} #1}{\mathrm{d}  #2}}
\renewcommand{\dfm}[3]{\frac{\mathrm{d} ^#1 #2}{\mathrm{d}  #3^#1}}
\renewcommand{\dft}[2]{\frac{\mathrm{d} ^2 #1}{\mathrm{d}  #2^2}}
\renewcommand{\md}{\mathrm{d}}
\renewcommand{\f}{p^\text{off}}
\renewcommand{\n}{p^\text{on}}
\renewcommand{\fw}{P^\text{off}}
\renewcommand{\nw}{P^\text{on}}
\renewcommand{\m}{p^\text{m}}
\renewcommand{\pd}{p^\text{d}}
\renewcommand{\mw}{P^\text{m}}
\renewcommand{\dw}{P^\text{d}}
\renewcommand{\ab}[1]{\langle #1 \rangle}
\renewcommand{\abp}[1]{\langle #1 \rangle_{\bs{p}}}
\renewcommand{\abpp}[1]{\langle #1 \rangle_{\bs{pp}}}
\renewcommand{\re}{\rho_\text{eq}}
\renewcommand{\R}{R_\text{eq}}
\renewcommand{\w}[1]{\widetilde{#1}}
\renewcommand{\pe}{\mathrm{Pe}}

\renewcommand{\nb}{\bs{\nabla}_{\bs{x}}}
\renewcommand{\np}{\bs{\nabla}_{\bs{p}}}
\renewcommand{\Dt}[1]{\frac{\mathrm{D} #1}{\mathrm{D} t}}

\renewcommand{\ol}[1]{\overline{#1}}

\newcommand{\Pe}{\text{Pe}}



\section{Continuum model} \label{sec::model}

In this section we derive a continuum model for the coupled dynamics of active matter and solvent flow. The model details are first published in our previous work \cite{yang24}, where a brief introduction and history of field-theory models on active networks are also included. We reformulate the model below in a more pedagogical manner for readers' convenience.

\subsection{Overview}

The main chemical components in our system are  microtubules, motors and ATP. The dynamics of active matter and solvent flows is initiated by a cascade of chemical reactions.  Firstly, the  motors are activated by light and reversibly linked. The linked motors then crosslink microtubules into a contracting network, which further drives a solvent flow. In our modeling, microtubules and motors are classified according to their chemical structures, such as unlinked or linked motors, and also to their crosslinking status, such as bound linked motors and free linked motors, as well as crosslinked and free microtubules. Here "bound" means  the motors are attached to the crosslinked microtubules, and "free" means  the microtubules or motors are not attached to the  crosslinked  network.

All the chemical reactions mentioned above, namely the reversible ``linking" of motors and the crosslinking of microtubules can be modeled in continuity equations of these chemical components. The light patterns enter our model through the dimerization rate $p^d(x,y,t)$, where $x$ and $y$ are the spatial coordinates and $t$ is time. The shape of the light region can thereby be  determined by a spatiotemporal function $p^d(x,y,t)$. 

\begin{figure}[h]
	\centering
	\includegraphics[width=0.7\linewidth,angle=0]{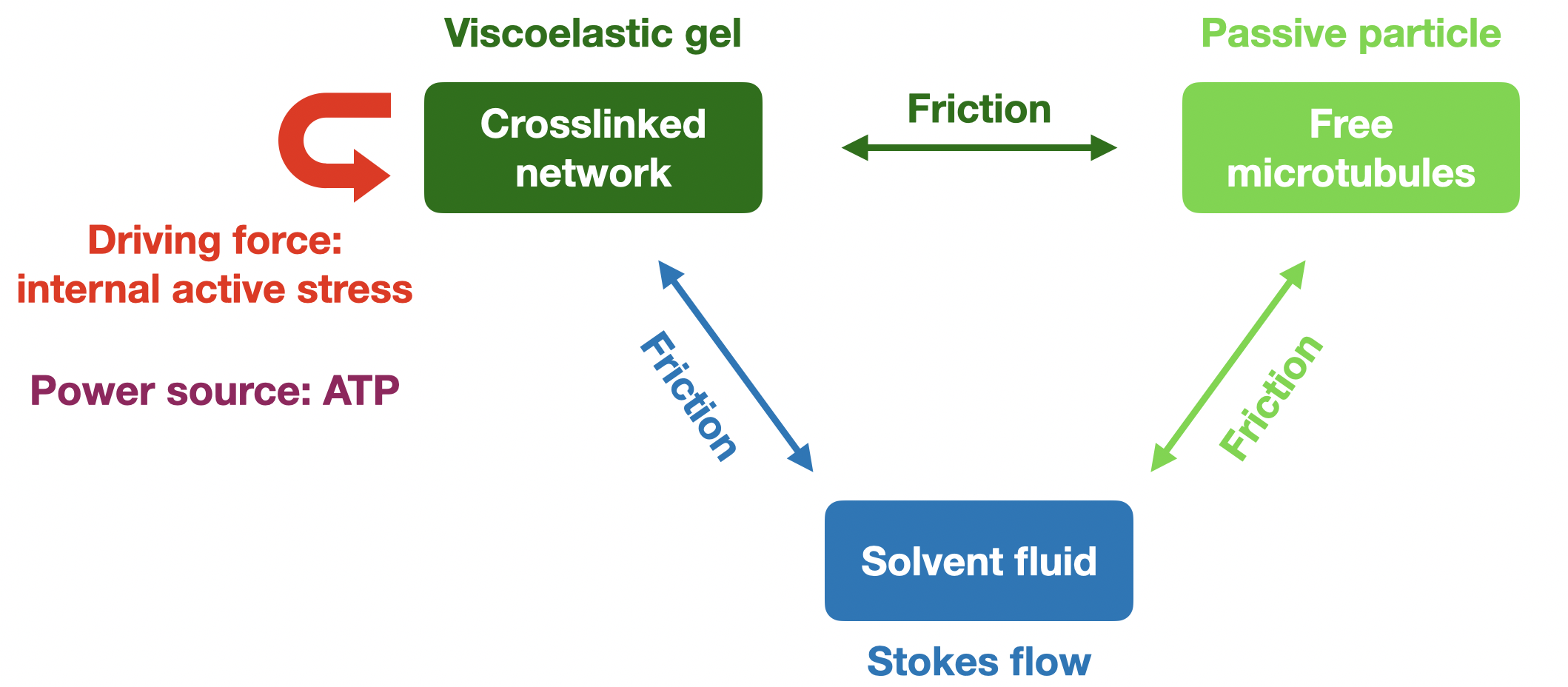}
	\caption{\textbf{Overview of the three-phase continuum model.}}
	\label{SMfig0}
\end{figure}

Momentum equations determine the motion of active matter and solvent from their stresses and material properties. In principle, we can write down momentum equations for each chemical species. However, small proteins such as free motors have little mechanical impact on the dynamics and can be assumed to follow the solvent flow. The crosslinked microtubules and bound linked motors can be modeled together as a single ``phase", which is termed  the   ``crosslinked network". The crosslinked network is modeled as a viscoelastic gel, which self-contracts driven by its internal active stresses; the solvent flow is modeled as a Stokes flow driven by the contraction of the active gel and balanced by the hydrodynamic resistance in the flow cell; the free microtubules are modeled as passive particles. Each of the three phases exert friction on the other two phases. A summary of the three phases is shown in Fig. \ref{SMfig0}.

\subsection{Continuity equations} \label{SM::subs::cont}

The continuity equations track the  change of mass of a component in time due to fluxes and chemical reactions. A general form of continuity equations can be written as
\begin{equation}
	\pf{c^*}{t}+ \on \cdot \bs{j}^* = \{ \mbox{chemical-reaction terms}\}, \label{SM::contgen}
\end{equation}
where  the asterisk  (*) is used to represent the general form of any chemical species in our system, $t$ is time, $c^*$ is  concentration,  and $\bs{j}^*$ is  flux. In our system, the flux $\bs{j}^*$ usually consist of two components, i.e.,
\begin{equation}
	\bs{j}^* =c^* \bs{v}^* - D^*\on c^*,
\end{equation}
where  $\bs{v}^*$ is velocity and $D^*$ is diffusivity. $c^* \bs{v}^*$ and $- D^*\on c^*$ are  called the convective and diffusive fluxes, respectively. We follow a convention to keep the convective flux on the left-hand side of (\ref{SM::contgen}) while moving  the diffusive flux to the right-hand side. Therefore,  the general form of continuity equations is
\begin{equation}
	\pf{c^*}{t}+ \on \cdot \left(c^* \bs{v}^*\right) = \{ \mbox{chemical-reaction terms}\} + D^*\nabla^2 c^*
\end{equation}

\begin{figure}[t]
	\centering
	\includegraphics[width=0.55\linewidth,angle=0]{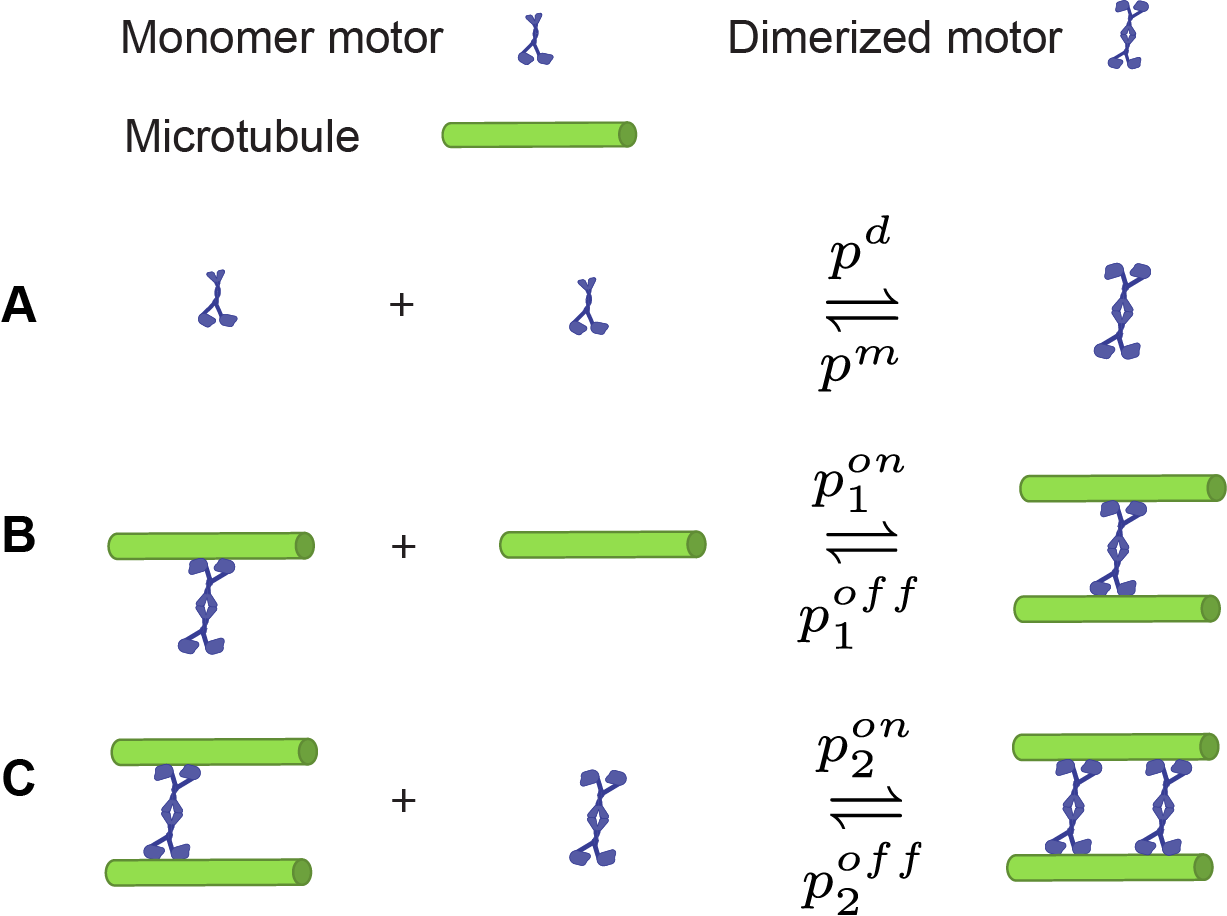}
	\caption{\textbf{Illustrations of chemical reactions in the model}. \textbf{A}, Reversible linking of motors. \textbf{B}, Reversible crosslinking of a freely-moving microtubule $i$ and another microtubule $j$. The latter can be either free or pre-crosslinked. \textbf{C}, Reversible binding of a linked motor on  two pre-crosslinked microtubules.}
	\label{SMfig1}
\end{figure}
The chemical reactions in our system are sketched in Fig. \ref{SMfig1}. The motor proteins can reversibly link and the continuity equation is
\begin{equation}
	\pf{m}{t} + \on \cdot (m \bs{u}) = 2 (p^m d_f - p^d m^2) + D_m \nabla^2 m \label{SM::m}
\end{equation}
where $m$ is the unlinked motor concentration, $d_f$ is the free linked motor concentration, $\bs{u}$ is the solvent flow velocity,  $D_m$ is the diffusivity,  $p^m$ and $p^d(x,y,t)$ are the monomerization and dimerization rates, respectively. The function $p^d(x,y,t)$ is determined by the light pattern. In experiments, we use V-shaped light patterns, which can be defined by a set of points $A_l = \{(x,y) \in \mbox{light pattern} \}$. The light pattern is projected periodically onto the flow cell with time interval $T$ and duration $\Delta t$. Naively, the dimerization rate can be written as
\begin{equation}
	p^d_i(x,y,t) = \begin{cases}
		p^d_0 & (x,y) \in A_l \ \mbox{and } t\in [nT, nT+\Delta t],\\
		0 &\text{otherwise},
	\end{cases} \label{SM::naivepd}
\end{equation}
where $p^d_0$ is a constant dimerization rate and $n=0,1,2,\dots$ is the number of  light pulses. However, directly using (\ref{SM::naivepd}) in simulations will cause numerical instability due to discontinuity of $p^d_i(x,y,t)$ in space. There are multiple ways to deal with this. In this Letter, we use a Gaussian kernel smoother to make (\ref{SM::naivepd}) a smooth function in space. The Gaussian kernel is defined as
\begin{equation}
	K_r(x,y,X,Y) = e^{-\frac{(x-X)^2+(y-Y)^2}{2r^2}} \label{SM::smoother}
\end{equation}
with a constant $r$, which is the ``kernel radius". Denoting the whole calculation domain as $A$, the smooth dimerization rate $p^d(x,y,t)$ can be obtained from applying the smoother (\ref{SM::smoother}) on (\ref{SM::naivepd}) through
\begin{equation}
	p^d(x,y,t) = \frac{\int_A p^d_i(X, Y,t)K_r(x,y,X,Y)\md X\md Y}{\int_A K_r(x,y,X,Y)\md X\md Y}.
\end{equation}
The slope of $p^d(x,y,t)$ at the boundary of light pattern $A_l$ can be tuned by the parameter $r$. A smaller $r$ yields a larger slope and thereby a sharper light boundary in simulations.

The continuity equations for the crosslinked ($c$) and free ($c_f$) microtubules are
\begin{subeqnarray}
	\pf{c}{t}+\on \cdot \left( c\bs{v}\right) = p^{\text{on}}_1 c_f d_f - p^{\text{off}}_1 c,\slabel{SM::cont_c}\\
	\pf{c_f}{t}+\on \cdot \left( c_f\bs{v}_f\right) = -p^{\text{on}}_1 c_f d_f + p^{\text{off}}_1  c + D_f\nabla^2 c_f. \label{SM::c}
\end{subeqnarray}
where  $D_f$ is the diffusivity of the free microtubules, $p^{\text{on}}_1$ and $p^{\text{on}}_1$ are the crosslinking and un-crosslinking rates, respectively, see Fig. \ref{SMfig1}B. The velocities of the crosslinked and free microtubules are denoted by $\bs{v}$ and $\bs{v}_f$, respectively. In (\ref{SM::cont_c}) we neglect the diffusion flux of the crosslinked microtubules. We also note that in writing down the chemical reaction terms in (\ref{SM::c}), we assume the reaction orders are 1 for simplicity.  

The continuity equations for the free ($d_f$) and the bound ($d_b$) linked motors are
\begin{subeqnarray}
	\pf{d_f}{t} + \on \cdot \left( d_f \bs{u}\right) = -p^m d_f + p^d m^2 - p_1^{\text{on}}c_fd_f-p_2^{\text{on}}cd_f+p^{\text{off}}_2d_b+D_d\nabla^2 d_f\\
	\pf{d_b}{t}+\on \cdot \left( d_b \bs{v}\right) =  p_1^{\text{on}}c_fd_f + p_2^{\text{on}}cd_f - p^{\text{off}}_2d_b, \label{SM::d}
\end{subeqnarray}
where $d_b$ is the bound linked motor concentration,  $p_2^{\text{on}}$ is the rate of the linked motors binding on the crosslinked microtubules, $D_d$ is the diffusivity of the free linked motors, $p^{\text{off}}_2$ is the unbinding rate of a linked motor that does not uncrosslink the two microtubules, see Fig. \ref{SMfig1}C.  All free motors are assumed to follow the solvent flow ($\bs{u}$) and the bound motors follow the crosslinked gel ($\bs{v}$) .

The continuity equation for the ATP is
\begin{equation}
	\pf{c_A}{t} + \on \cdot(c_A \bs{u}) = -k_Ad_bc_A +D_A\nabla^2c_A, \label{SM::ATPd}
\end{equation}
where $k_A$ is the consumption rate of the ATP by the bound linked motors, and $D_A$ is the ATP diffusivity.

The solvent flow is incompressible, meaning that its density does not change. The continuity equation for the solvent flow is just
\begin{equation}
	\on \cdot \bs{u} = 0. \label{SM::contu}
\end{equation}

\subsection{Momentum equations}  \label{SM::subs::mom}

Momentum equations are the balance between the momentum fluxes, stresses and body forces in a differential form. At low Reynolds number, inertial momentum fluxes are usually neglected. Under this assumption, a general form of the momentum equation is
\begin{equation}
	\on \cdot \bs{\sigma}^* + \bs{f}^* = \bs{0}, 
\end{equation}
where $ \bs{\sigma}^* $ is the total stress and $\bs{f}^* $ is the total body force. In this section, we list the momentum equations for each phase. 

\subsubsection{Crosslinked network as an active gel}

The crosslinked network forms an active gel, which is assumed to be viscoelastic. The momentum equation can be written as
\begin{equation}
	\on \cdot \left(\bs{\sigma}_a + \bs{\sigma}_v+ \bs{\sigma}_{el} + \bs{\sigma}_{st}\right) + \bs{f}_{fl} + \bs{f}_f = \bs{0}, \label{SM::moma}
\end{equation}
where $\bs{\sigma}_a$, $\bs{\sigma}_v$,  $\bs{\sigma}_{el}$ and $\bs{\sigma}_{st}$ are the active, viscous, elastic and steric stresses in the gel, $\bs{f}_{fl}$ and $\bs{f}_f$ are the friction with the solvent and freely-moving microtubules, respectively. Their explicit expressions can be derived from microscopic interactions \cite{furthauer19,furthauer21}. The active contractile stress is assumed to be
\begin{equation}
	\bs{\sigma}_a = \alpha c_A d_b c \bs{I}, \label{SM::sigmaa}
\end{equation}
where $c$, $c_A$ and  $d_b$ are the concentrations of the crosslinked microtubules, ATP and the bound linked motors, respectively, $\alpha$ is a constant activity coefficient and $\bs{I}$ is the identity tensor. This active stress is derived to be $\bs{\sigma}'_a \propto c^2 \bs{I},$ in Ref \cite{furthauer21}, where the bound motor concentration is implicitly assumed to be uniform and does not appear in active stresses. In our experiments, the concentration difference of the linked motors inside and outside of the illuminated regions is the key to form active gels. We can modify the theory in Ref \cite{furthauer21} by adding a prefactor $d_b/c$ in the active stresses, which is the number of bound motors per microtubule. 

The viscous stress is 
\begin{equation}
	\bs{\sigma}_v=\eta d_b c \left( \on \bs{v} + \on \bs{v} ^T\right), \label{SM::sigmav}
\end{equation}
where $\bs{v}$ is the velocity of the crosslinked microtubules and $\eta$ is the viscosity coefficient. This viscous stress in Ref \cite{furthauer21} is $\bs{\sigma}'_v\propto c^2 \left( \on \bs{v} + \on \bs{v} ^T\right)$. In our model the coefficient is modified from $c^2$ to $d_bc$ to incorporate the effects of the motor concentrations. 

The elastic stress $\bs{\sigma}_{el}$ is assumed to follow the Oldroyd-Maxwell model with a long relaxation time \cite{alves21},
\begin{equation}
	\Dt{\bs{\sigma}_{el}}-\left[ \bs{\sigma}_{el} \cdot \on \bs{v} + \on \bs{v}^T \cdot \bs{\sigma}_{el}  \right] = \eta_{el} d_b c \left( \on \bs{v} + \on \bs{v} ^T\right) - p^{\text{off}}_1 \bs{\sigma}_{el},
\end{equation}
where  $\eta_{el}$ is the elasticity coefficient and $\mathrm{D}/\mathrm{D}t$ is the material derivative. We choose the Oldroyd-Maxwell model for its simplicity.

To avoid unlimited density build-up at one point, we add a steric stress that is 
\begin{equation}
	\bs{\sigma}_{st} =- \xi c^2 \bs{I},
\end{equation}
with a constant coefficient $\xi$.

The friction with freely-moving microtubules is 
\begin{equation}
	\bs{f}_f = \beta cc_f(\bs{v}_f-\bs{v}),
\end{equation} 
where $\beta$ is the friction coefficient. The friction with the solvent flow is 
\begin{equation}
	\bs{f}_{fl} = \gamma c(\bs{u}-\bs{v}) \label{SM::fl}
\end{equation}
with the drag coefficient $\gamma$. In general the hydrodynamic drag coefficient depends on the fiber orientation \cite{leal07}. In this Letter we neglect the polarity in both active stresses and  hydrodynamic friction.

\subsubsection{Free microtubules}
The free microtubules are passive particles that only experience friction from the gel and the ambient fluid. The force balance is $\beta c_fc\left(\bs{v}-\bs{v}_f \right) + \gamma c_f (\bs{u}-\bs{v}_f)=\bs{0}$, and their velocity is therefore 
\begin{equation}
	\bs{v}_f = \frac{\beta c\bs{v} + \gamma \bs{u}} { \beta c + \gamma}. \label{SM::vf}
\end{equation}

\subsubsection{Solvent flows}

The flow cells used in our experiments are usually called  Hele-Shaw cells in fluid mechanics,  meaning that their horizontal dimensions ($x-$ and $y-$direction) greatly exceed their vertical dimension ($z$-direction). We now derive the averaged two-dimensional flow equation in the $xy-$plane. 
We start from the three-dimensional Stokes equation, that is,
\begin{equation}
	-\tn \w{\Pi} +\mu \w{\nabla}^2 \wb{u}+\gamma \w{c}  \left(\wb{v}-\wb{u}\right) + \gamma \w{c}_f  \left(\wb{v}_f-\wb{u}\right)=0, \label{SM::stokes}
\end{equation}
where $\w{\Pi}$ and $\mu$ are the fluid pressure and viscosity, respectively. We use a tilde ``$\sim$" on top to indicate the variable is a function of $(x, y, z)$ and if the variable is a vector, it is a three-dimensional vector, e.g., $\w{\Pi} = \w{\Pi}(x,y,z)$ and $\wb{u}(x,y,z) = \left(\w{u}_x(x,y,z), \w{u}_y(x,y,z), \w{u}_z(x,y,z)\right)$. Since the vertical length scale is much smaller than horizontal length scales, the classical lubrication theory \cite{leal07} shows that $\w{\Pi}(x, y, z) \approx \Pi(x, y)$, i.e., the fluid pressure is constant along the $z$-direction, and $\w{u}_z\approx 0$. Substituting these two results into equation (\ref{SM::stokes}) yields $\w{v}_z\approx 0$ and $\w{v}_{f,z}\approx 0$. Furthermore, the $x-$ and $y-$components of equation (\ref{SM::stokes}) can be approximated by 
\begin{equation}
	-\on \Pi +\mu \pft{\wb{u}}{z} +\gamma \w{c}  \left(\wb{v}-\wb{u}\right) + \gamma\w{c}_f  \left(\wb{v}_f-\wb{u}\right)=0, \label{SM::3dstokes}
\end{equation}
where all vectors only have $x-$ and $y-$components, i.e., $\wb{u} = \left(\w{u}_x(x,y,z), \w{u}_y(x,y,z)\right)$,  $\wb{v} = \left(\w{v}_x(x,y,z), \w{v}_y(x,y,z)\right)$ and  $\wb{v}_f = \left(\w{v}_{f,x}(x,y,z), \w{v}_{f,y}(x,y,z)\right)$. Therefore, the analytical solution of $\wb{u}$ requires knowledge of the three-dimensional distribution of $\w{c}$, $\wb{v}$ and $\wb{v}_f$. To derive a two-dimensional model, we further simplify equation (\ref{SM::3dstokes}) by replacing the friction with their $z$-directional average, i.e.,
\begin{equation}
	-\on \Pi +\mu \pft{\wb{u}}{z} +\gamma c  \left(\bs{v}-\bs{u}\right) + \gamma c_f  \left(\bs{v}_f-\bs{u}\right)=0, \label{SM::2dstokes}
\end{equation}
where $c(x,y)=h^{-1}\int_{0}^{h}\w{c}\md z$, $\bs{v}(x,y)=h^{-1}\int_{0}^{h}\wb{v}\md z$, $\bs{v}_f(x,y)=h^{-1}\int_{0}^{h}\wb{v}_f\md z$, and $\bs{u}(x,y)=h^{-1}\int_{0}^{h}\wb{u}\md z$ with the top and bottom walls of channel at $z=0$ and $z=h$, respectively. These notations are consistent with the rest of the Letter. The solution of equation (\ref{SM::2dstokes}) is
\begin{equation}
	\wb{u}=\frac{1}{2\mu}\left[\on \Pi -\gamma c\left(\bs{v}-\bs{u}\right) - \gamma c_f \left(\bs{v}_f-\bs{u}\right)\right]z(z-h). \label{SM::uf}
\end{equation}
Combining $\bs{u}=h^{-1}\int_{0}^{h}\wb{u}\md z$ and equations (\ref{SM::vf}) and (\ref{SM::uf}), we have the equation for the averaged two-dimensional flow velocity
\begin{equation}
	\bs{u}=-\frac{h^2}{12\mu}\left[\on \Pi -\left(\gamma c + \gamma_f c_f\right)\left(\bs{v}-\bs{u}\right) \right], \label{SM::u}
\end{equation}
with $\gamma_f = \gamma \beta c /(\beta c+\gamma)$. The above result is a modified Darcy's law in a Hele-Shaw cell \cite{leal07} by incorporating the microtubule friction.

\subsection{Non-dimensionalization}

Using the initial microtubule concentration $c_0$, initial unlinked motor concentration $m_0$, initial ATP concentration $c_{A0}$, and the typical length scale of the illuminated region $l$, we can non-dimensionalize the governing equations by
\begin{equation}
	\begin{split}
		\ol{c}=\frac{c}{c_0}, \quad \obn= l_0 \on, \quad \ol{m}=\frac{m}{m_0}, \quad \ol{d}_f=\frac{d_f}{m_0}, \quad \ol{d}_b=\frac{d_b}{m_0}, \quad \ol{c}_A=\frac{c_A}{c_{A0}} \\
		\ol{t}=\frac{t}{t_0}, \quad  \ob{v}=\frac{\bs{v}}{v_0} , \quad  \ob{v}_f=\frac{\bs{v}_f}{v_0},  \quad \ob{u}=\frac{\bs{u}}{v_0},\quad \ob{\sigma}_{el} = \frac{\bs{\sigma}_{el}}{\sigma_0},\quad \ol{\Pi}=\frac{\Pi}{\Pi_0},
	\end{split}
\end{equation}
where we use  overlines to denote dimensionless variables, $t_0$, $v_0$, $\sigma_0$ and $\Pi_0$ are  typical scales of time, velocity,  stress, and fluid pressure, respectively. By balancing the contractile stress (\ref{SM::sigmaa}) and the viscous stress (\ref{SM::sigmav}) in the active gel, we have $v_0 = \alpha c_{A0}l/\eta$. The  typical time scale follows as $t_0=l_0/v_0 = \eta /\alpha c_{A0}$. The typical stress scale can be obtained from equation (\ref{SM::sigmaa}), which is $\sigma_0 = \alpha c_{A0}c_0m_0$. From equation (\ref{SM::u}), we have $\Pi_0 = c_0\gamma v_0 l_0 =\alpha \gamma c_0 c_{A0} l_0^2/\eta $.

The dimensionless continuity equations for the unlinked (\ref{SM::m}) and linked (\ref{SM::d}) motors are
\begin{subeqnarray}
	\pf{\ol{m}}{\ol{t}} + \obn \cdot (\ol{m} \ob{u}) = 2 (\ol{p}^m \ol{d}_f - \ol{p}^d \ol{m}^2) + \Pe_m^{-1} \ol{\nabla}^2 \ol{m}, \\
	\pf{\ol{d}_f}{\ol{t}} + \obn \cdot \left( \ol{d}_f \ob{u}\right) = -\ol{p}^m \ol{d}_f + \ol{p}^d \ol{m}^2 - \ol{p}_1^{\text{on}}\ol{c}_f\ol{d}_f-\ol{p}_2^{\text{on}}\ol{c}\ol{d}_f+\ol{p}^{\text{off}}_2\ol{d}_b+\Pe_d^{-1} \ol{\nabla}^2 \ol{d}_f,\\
	\pf{\ol{d}_b}{\ol{t}}+\obn \cdot \left( \ol{d}_b \ob{v}\right) =  \ol{p}_1^{\text{on}}\ol{c}_f\ol{d}_f + \ol{p}_2^{\text{on}}\ol{c}\ol{d}_f - \ol{p}^{\text{off}}_2\ol{d}_b, 
\end{subeqnarray}
where the dimensionless reaction coefficients are $\ol{p}^m =p^m t_0$, $\ol{p}^d =p^m m_0 t_0$, $\ol{p}^{\text{on}}_1 =p^{\text{on}}_1 c_0 t_0$, $\ol{p}^{\text{on}}_2 =p^{\text{on}}_2 c_0 t_0$ and $\ol{p}^{\text{off}}_2 =p^{\text{off}}_2 t_0$. We use $\Pe$ to denote the P\'{e}clet number and $\Pe_m = v_0 l /D_m$, $\Pe_d = v_0 l /D_d$.

The dimensionless continuity equations for the microtubules (\ref{SM::c}) are
\begin{subeqnarray}
	\pf{\ol{c}}{\ol{t}}+\obn \cdot \left( \ol{c}\ob{v}\right) = \ol{p}^{\text{on}}_1 \ol{c}_f \ol{d}_f - \ol{p}^{\text{off}}_1 \ol{c},\\
	\pf{\ol{c}_f}{\ol{t}}+\obn \cdot \left( \ol{c}_f\ob{v}_f\right) = -\ol{p}^{\text{on}}_1 \ol{c}_f \ol{d}_f + \ol{p}^{\text{off}}_1  \ol{c} + \Pe_f^{-1}\ol{\nabla}^2 \ol{c}_f \label{po1}
\end{subeqnarray}
with $\ol{p}^{\text{off}}_1 = p^{\text{off}}_1 t_0 $, $\Pe_f = v_0 l_0 /D_f$.  For the ATP, equation (\ref{SM::ATPd}) becomes
\begin{equation}
	\pf{\ol{c}_A}{\ol{t}}+\obn \cdot \left( \ol{c}\ob{u}\right) = -\ol{k}_A \ol{d}_b \ol{c}_A + Pe_A^{-1}\ol{\nabla}^2 \ol{c}_A,
\end{equation}
where $\ol{k}_A=k_A m_0 t_0$ and $\Pe_A = v_0 l_0 /D_A$.

The continuity equation for the solvent flow (\ref{SM::contu}) is
\begin{equation}
	\obn \cdot \ob{u}=0.
\end{equation}

Using equations (\ref{SM::moma}-\ref{SM::vf}), the dimensionless momentum equations for the active gel are
\begin{equation}
	\obn \cdot \left[\left(\ol{c}_A\ol{d}_b\ol{c} -\ol{\xi}\ol{c}^2 \right)\bs{I}+\obn \ob{v}+\obn \ob{v} ^T+ \ob{\sigma}_{el}\right] + \left(\ol{\gamma}_f\ol{c}_f + \ol{\gamma}\ol{c}\right)\left(\ob{u}-\ob{v}\right) = \bs{0}
\end{equation}
where $\ol{\xi} = \xi c_0 /\alpha c_{A0}m_0$, $\ol{\gamma}=\gamma l_0^2/m_0 \eta$,  $\ol{\gamma}_f=\gamma_f l_0^2/m_0 \eta$ and
\begin{equation}
	\frac{\mathrm{D} \ob{\sigma}_{el}}{\mathrm{D} \ol{t}}-\left[ \ob{\sigma}_{el} \cdot \obn \ob{v} + \obn \ob{v}^T \cdot \ob{\sigma}_{el}  \right] = \ol{\eta}_{el} \ol{d}_b \ol{c} \left( \obn \ob{v} + \obn \ob{v} ^T\right) - \ol{p}^{\text{off}}_1 \ob{\sigma}_{el},
\end{equation}
with $\ol{\eta}_{el} = \eta_{el}/\alpha c_{A0}$.

The dimensionless velocity of the freely-moving microtubules (\ref{SM::vf}) is 
\begin{equation}
	\ob{v}_f =\frac{\ol{\beta}\ol{c}\ob{v}+\ol{\gamma}\ob{u}}{\ol{\beta}\ol{c}+\ol{\gamma}},
\end{equation}
with $\ol{\beta}=\beta l_0^2 c_0 / m_0 \eta$. Note that only two of $\ol{\beta}$, $\ol{\gamma}$ and $\ol{\gamma}_f$ are independent, which are connected through $\ol{\gamma}_f =\ol{\gamma} \ol{\beta} \ol{c}/ (\ol{\beta} \ol{c} + \ol{\gamma})$.

The dimensionless solvent flow (\ref{SM::u}) is
\begin{equation}
	\ob{u} = -\ol{\zeta}\left[ \obn \ol{\Pi}-\left(\ol{c}+\frac{\ol{\gamma}_f}{\ol{\gamma}}\ol{c}_f\right)\left(\ob{v}-\ob{u}\right)\right],
\end{equation}
with $\ol{\zeta} = h^2c_0 \gamma /12\mu$.

We used the finite difference method in numerical simulations with the central difference scheme in space and the method of lines in time.   
The codes are written in Python and available at
https://github.com/fy26/ActiveHealing.

The parameters  used in simulations are
\begin{eqnarray}
	l_0 = 220 \ \mu \mbox{m}, \quad t_0 = 40 \ \mbox{s}, \quad p^d_0 = 15, \quad p^m = 12, \quad p^{\text{on}}_1 = 15, \quad p^{\text{on}}_2 = 0.75, \nonumber \\
	p^{\text{off}}_1 = 10, \quad   p^{\text{off}}_2 = 20, \quad \Pe_f^{-1} =10^{-3}, \quad \Pe_m^{-1} = \Pe_d^{-1} = \Pe_A^{-1} = 10^{-2}, \nonumber\\
	\ol{k}_A = 0.8,  \quad \ol{\xi}= 0.08, \quad \ol{\eta}_{el} = 1.5, \quad \ol{\gamma} = 0.045, \quad \ol{\beta}=0.45, \quad \ol{\zeta} = 20.5.
\end{eqnarray}
In addition, the duration of each light pulse, $\Delta t / t_0 = 0.04$ and the time interval between two light pulses is $T / t_0 = 0.2$.
\subsection{Simulations}

\subsubsection{Bifurcation diagram with varying  crosslinking rates}

\begin{figure}[h]
	\includegraphics[width=0.6\linewidth,angle=0]{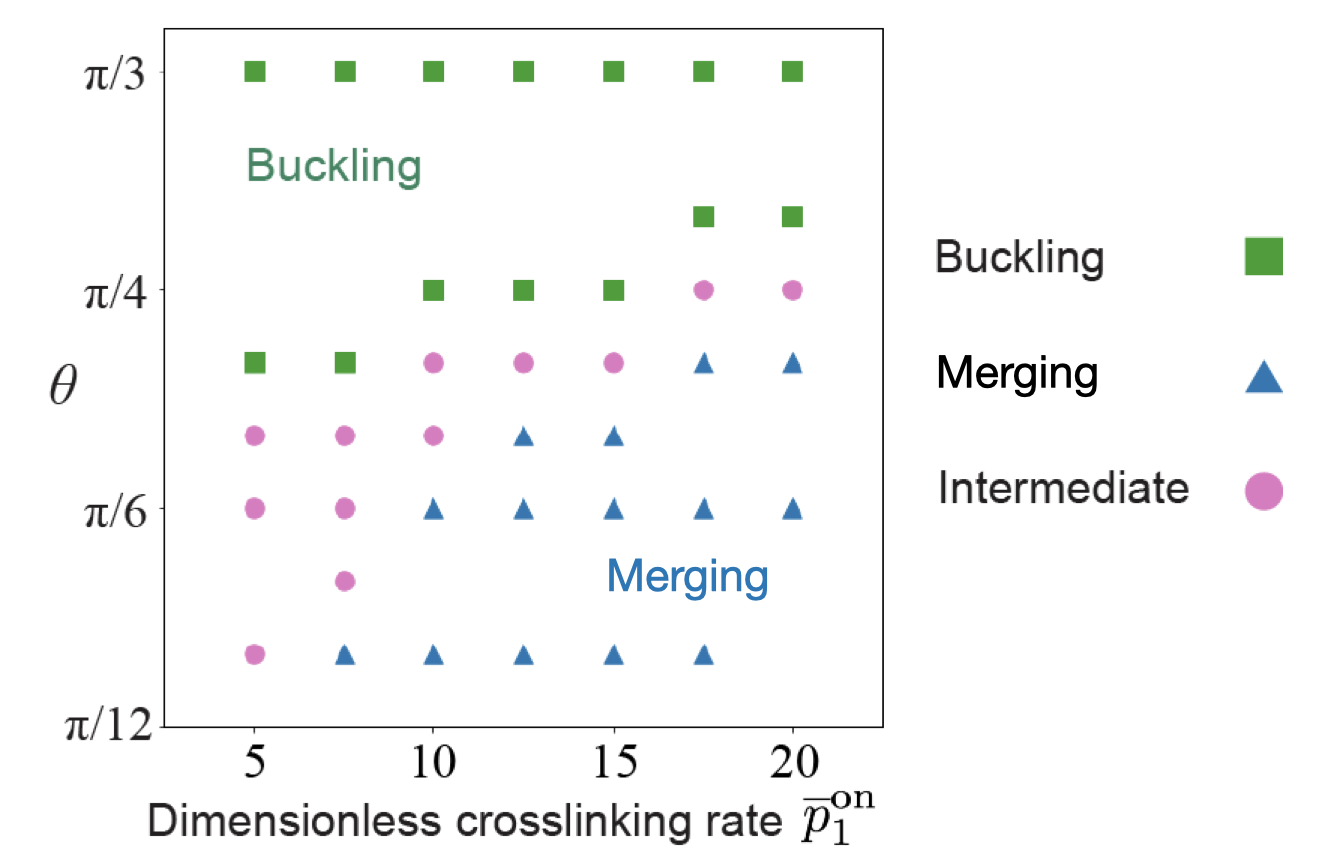}
	\caption{Simulated bifurcation diagram shows that the critical angle increases with the crosslinking rate. The diagram documents whether the network in Fig. 2(d) merges or buckles with varying crosslinking rates and initial crack angles. ``Intermediate" represents when the network does not buckle and also does not show significant merging, such as Fig. 2(d). In simulations,  the ``Intermediate"  state is characterized by the average curvature close to 0.} 
	\label{SM::p1}
\end{figure}

To demonstrate that the merging is driven by active crosslinking, we simulate the bifurcation diagram of the network in Fig. 2(d) in the main text by varying the crosslinking rates $\ol{p}^{\text{on}}_1$ in (\ref{po1}) and documenting whether the network buckles or merges under different initial crack angles $\theta$. The result is  in Fig. \ref{SM::p1}, showing that increasing the crosslinking rate promotes network merging, thereby confirming that the merging process is driven by active crosslinking.

\subsubsection{Boundary layer of free microtubules and motors}

\begin{figure}[t]
	\includegraphics[width=1\linewidth,angle=0]{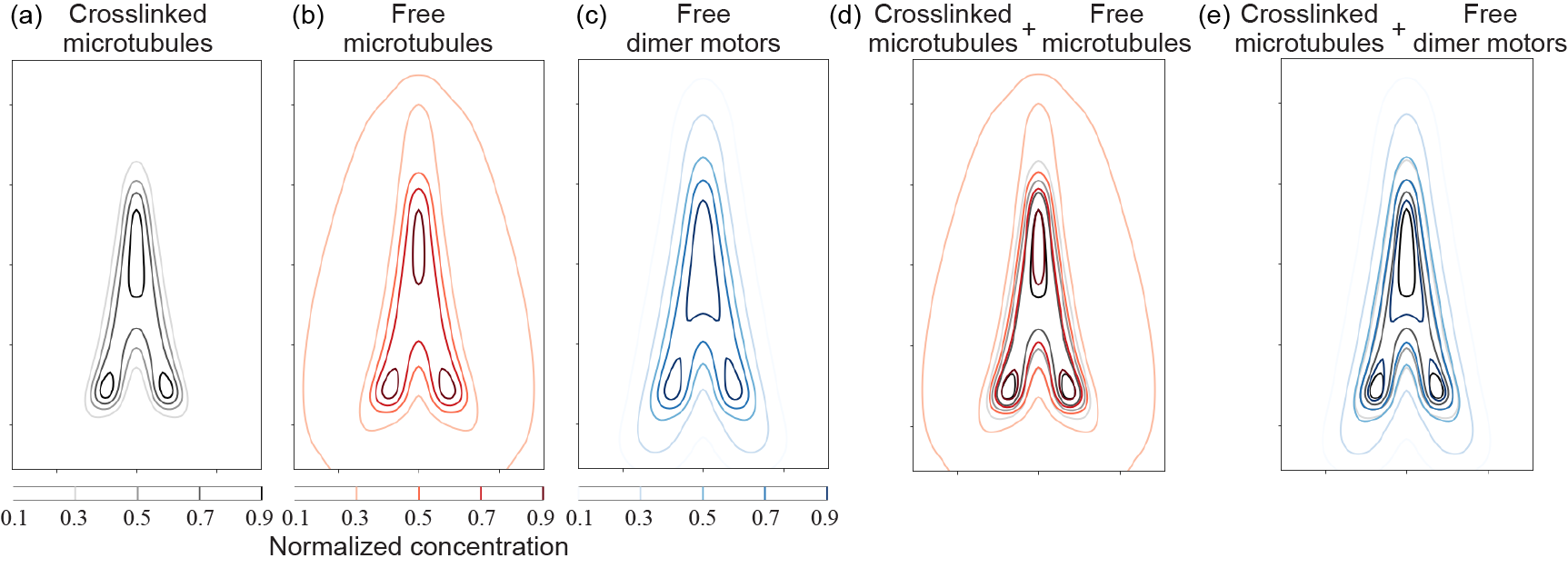}
	\caption{ Normalized concentration contours of crosslinked microtubules, free microtubules and free linked motors. The contours of crosslinked microtubules (a) are encapsulated by the contours  of free microtubules  (b) and  free linked motors (c) at the same levels, demonstrating that there is a boundary layer of free microtubules and motors outside the network surface.  (e) and (f) are overlaid contours of crosslinked and free microtubules, and crosslinked and free linked motors, respectively.} \label{SM::BL}
\end{figure}

We plot normalized concentrations of crosslinked microtubules, free microtubules and free linked motors in Fig. \ref{SM::BL}. The contours are simulated results of Fig 2(b) in the main text at $\overline{t}=3$. From Fig. \ref{SM::BL} we can see that the contours of crosslinked microtubules are always encapsulated by the contours of free microtubules and motors, confirming that there is a boundary layer at the surface of active networks.

\subsection{ Force balance in the elastic-rod model.}

\begin{figure}[h]
	\centering
	\includegraphics[width=0.9\linewidth,angle=0]{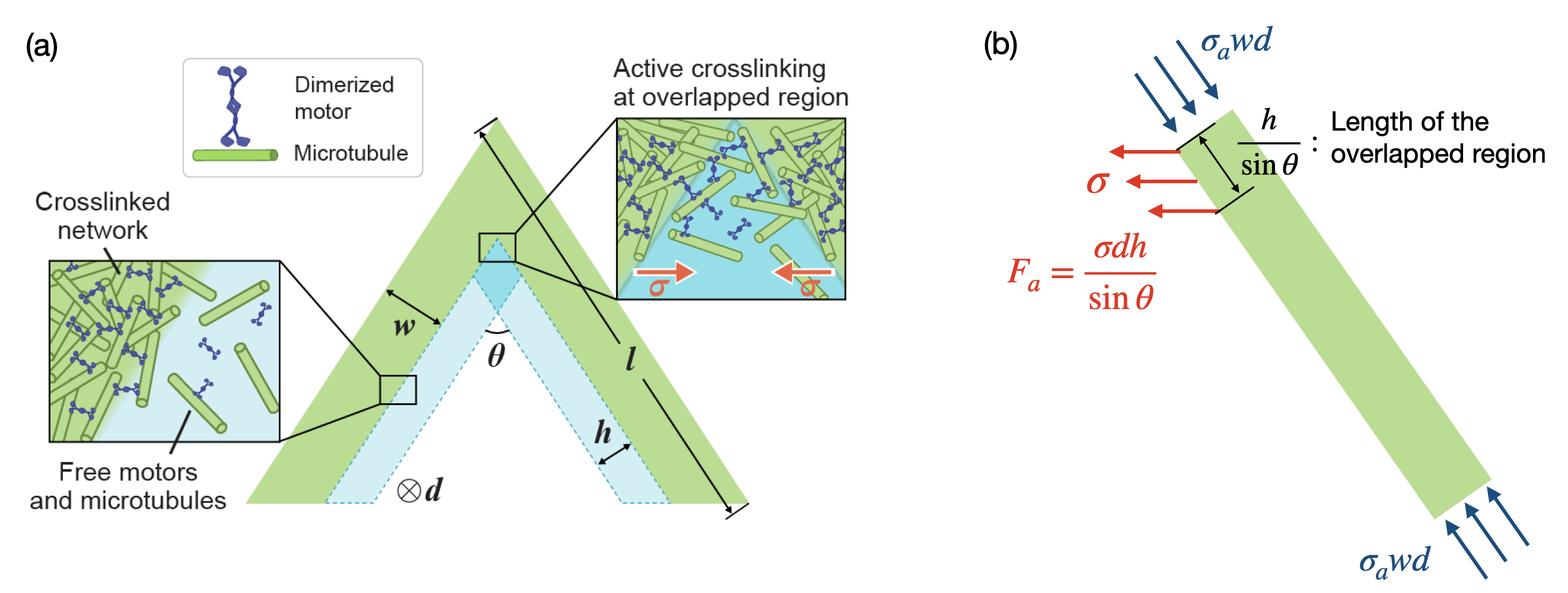}
	\caption{(a) Schematics of the elastic-rod model, identical to Fig. 4 in the main text. (b) Schematics of the active healing force $F_a = \sigma d h/ \sin \theta$ and the active compression $\sigma_a w d$ on the right arm. The viscous force exerted by the background solution on the arm is not plotted.}
	\label{Fig::force}
\end{figure}

To illustrate how forces act on each arm in the elastic-rod model, we depict the active healing force $F_h = \sigma d h /\sin\theta$ and the active compression $\sigma_a wd$ in Fig. \ref{Fig::force}. 

\subsection{Motor activity and active healing}

\begin{figure}
	\includegraphics[width=0.7\linewidth,angle=0]{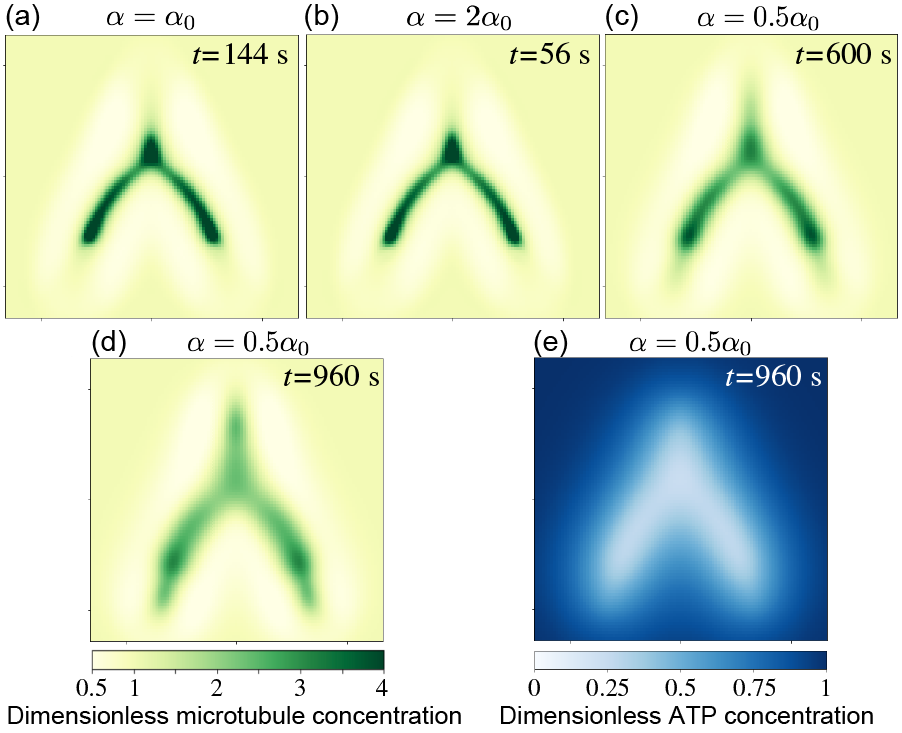}
	\caption{Simulations with different activities $\alpha$ show that the activity does not affect the arm curvature but can change the time scales and densities of the active networks. Simulation in (a) is the same as Fig. 2(c) in the main text and used as a benchmark. Increasing (b) or decreasing (c) the activity $\alpha$ can significantly speed up or slow down the active matter dynamics, respectively. (d) At late times, smaller activity yields a loosely crosslinked network. (e)  Depletion of ATP corresponding to (d). The ATP concentration is non-dimensionalized by the initial ATP concentration. }
	\label{activity}
\end{figure}

How does activity influence the healing process? The active stress $\bs{\sigma}_a$ in our simulations is modeled as an isotropic contractile stress $\sigma_a = \alpha c_A d_b c \bs{I}$, where $\alpha$ is the activity, $c_A$ is the ATP concentration, and $d_b$ is the concentration of motors bound to the active network. We numerically study how the activity $\alpha$ affects the bifurcation dynamics. Using the simulation in Fig. 2(c)  in the main text as a benchmark, we find that decreasing activity can slow down the active matter dynamics (Fig. \ref{activity}). The smaller activity also yields a more loosely crosslinked network at late times, as shown in Fig. \ref{activity}(d), in contrast with the dense networks in Fig. 2(c) in the main text. However, the activity does not affect the arm curvature, as seen from Fig. \ref{activity}(a-c). We also vary the activity on networks with different opening angles and get the same conclusions. The unvarying curvatures indicate that changing activity does not impact the competition of the merging and buckling forces and therefore does not change the critical angle $\theta^*$ in bifurcation. This is not surprising as both the merging and compression forces originate from the same microscopic active stress $\bs{\sigma}_a$.  Our simulations suggest that increasing activity can enhance both healing and compression forces in a similar manner such that even though the overall dynamics is accelerated, the critical angle $\theta^*$, which depends on the ratio of the compression and merging forces, remains largely unaffected. The loosely crosslinked network in Fig. \ref{activity}(d) is a natural consequence of weaker contractile stresses. Additionally,  the energy consumption, or equivalently, the ATP consumption also plays an important role in determining the late-stage network densities. In our simulations, the ATP consumption rate is modeled as $- k_Ad_bc_A$ with a constant coefficient $k_A$ (see equation (S10)). Note that we assume $k_A$ is unrelated to the activity $\alpha$. Since a smaller activity prolongs the dynamics and consumes more ATP over a longer time span, at late times the depletion of ATP, as shown in Fig. \ref{activity}(e), will stall further contraction of the network, which also gives rise to a more loosely crosslinked network. 

\section{Experimental setup} \label{sec::exp}
\subsection{Light-induced kinesin expression and purification}
We constructed two chimeras of D. melanogaster kinesin K401: K401-iLID and K401-micro as previously described (Addgene 122484 and 122485) \cite{ross19}. For the K401-iLiD plasmid we inserted iLID with a His tag after the C-terminus of K401. For the K401-micro plasmid, we inserted K401 between the His-MBP and the micro. The MBP domain is needed to ensure the microdomain remains fully functional during expression \cite{Guntas2015-to}. After the expression, the MBP domain can be cleaved off by TEV protease.

For protein expression, we transformed the plasmids to BL21(DE3)pLysS cells. The cells were grown in LB and induced at with 1mM IPTG at 18 $^\circ$C for 16 hours after reaching OD 0.6. The cells were then pelleted at 4000G and resuspended in lysis buffer (50 mM sodium phosphate, 4 mM MgCl$_2$, 250 mM NaCl, 25 mM imidazole, 0.05mM MgATP, 5 mM BME, 1 mg/ml lysozyme and 1 tablet/50 mL of Complete Protease Inhibitor). After a 1-hour incubation with stirring, the lysate was passed through a 30kPSI cell disruptor. The lysate was clarified at 30,000 G for 1 hour. The supernatant was then incubated with Ni-NTA agarose resin for 1 hour. The lysate/Ni-NTA mixture was loaded into a chromatography column and washed three times with wash buffer (Lysis buffer with no lysozyme nor complete EDTA tablet), and eluted with 500mM imidazole. Protein elutions were dialyzed overnight using 30 kDa MWCO membrane against 50 mM sodium phosphase, 4 mM MgCl$_2$, 250 mM NaCl, 0.05 mM MgATP, and 1 mM BME. For the K401-micro elution, we added TEV protease at a 1:25 mass ratio to remove the MBP domain. Then, we used centrifugal filters to exchange to pH 6.7 protein storage buffer (50 mM imidazole, HCl for pH balancing, 4 mM MgCl$_2$, 2 mM DTT, 50 $\mu$M MgATP, and 36\% sucrose). Proteins were then aliquoted and flash frozen in LN2 and stored under -80 $^\circ$C.

\subsection{Microtubule polymerization and length distribution}

Fluorescent microtubule polymerization was previously described \cite{ross19}. In short, we used a protocol based on one found on the Dogic lab homepage. The procedure began by preloading and starting a 37 $^\circ$C water bath. GMP-cpp, reagents, and tubes were cooled on ice. A 20 mM DTT solution was prepared using Pierce no-weigh format, and a GMP mixture consisting of M2B, DTT, and GMP-cpp was made and stored on ice. The ultracentrifuge and rotor were pre-cooled to 4 $^\circ$C. Tubulin (20 mg/mL) and labeled tubulin (20 mg/mL) were thawed in the water bath until mostly thawed and then cooled on ice. In a cold room, labeled tubulin was added to the stock vial of unlabeled tubulin and mixed gently. The GMP mixture was then added to the tubulin mixture and stirred gently. This combined mixture was pipetted into ultracentrifuge tubes and incubated on ice for 5 minutes before being centrifuged at 90,000 rpm, 4 $^\circ$C for 8 minutes. The supernatant was carefully collected without disturbing the pellet and transferred to an Eppendorf tube, mixed, and stored on ice. The mixture was incubated in a 37 $^\circ$C water bath for 1 hour, protected from light. Aliquots were then dispensed into PCR strip tubes, which were spun to collect the fluid at the bottom. Finally, the PCR strips were flash frozen in liquid nitrogen and stored in a -80 $^\circ$C freezer.

To measure the length distribution of microtubules, we imaged fluorescently labeled microtubules immobilized onto the cover glass surface of a flow cell. The cover glass was treated with a 0.01\% solution of poly-L-lysine (Sigma P4707) to promote microtubule binding. The lengths of microtubules were determined by image segmentation. Each microtubule image was normalized and underwent local and global thresholding to correct for non-uniform backgrounds and obtain thresholded images of putative microtubules. Morphological operations were applied to reconnect small breaks in filaments. Objects near the image boundary were removed, and small or circular objects were filtered out based on size and eccentricity thresholds. Potential microtubule crossovers were identified and eliminated by analyzing the angles of lines within the image.

\subsection{Sample chambers for experiments}
Glass slides and coverslips were first cleaned using a series of washes. Slides and coverslips were placed in respective containers, and 2\% Hellmanex solution was prepared by mixing 6 mL Hellmanex with 300 mL DI water, heated, and poured into the containers. The containers were sonicated for 10 minutes, followed by three DI water rinses and an ethanol rinse. Ethanol was added to the containers and sonicated again for 10 minutes, followed by another ethanol rinse and three DI water rinses. Next, 0.1 M KOH was added to the containers, sonicated for 10 minutes, and rinsed three times with DI water. The slides were then etched overnight with 5\% HCl and rinsed three times with DI water. Clean slides and coverslips were stored in DI water.

For silane coupling, a 2\% acrylamide solution was prepared using 40\% acrylamide stock solution, and degassed under vacuum. In a chemical hood, 98.5\% ethanol, 1\% acetic acid, and 0.5\% silane agent were mixed to prepare the silane-coupling solution, which was immediately poured into the containers with the slides and coverslips and incubated at room temperature for 20-30 minutes. The slides were then rinsed once with ethanol, three times with DI water, and baked at 110$^\circ$C for 30 minutes or 50$^\circ$C overnight.

For acrylamide polymerization, the degassed 300 mL of 2\% acrylamide solution was moved to a stir plate, and 105 $\mu$L TEMED and 210 mg ammonium persulfate were added. The solution was immediately poured over the silane-coupled slides and coverslips and left to polymerize overnight at 4$^\circ$C. Before use, the slides and coverslips were rinsed with DI water and air-dried.

\subsection{Reaction mixture}
For the self-organization experiments, K401-micro, K401-iLID, and microtubules were combined into a reaction mixture to achieve final concentrations of approximately 0.1 $\mu$M for each motor type and 1.5-2.5 $\mu$M for tubulin, referring to protein monomers for K401-micro and K401-iLID constructs, and protein dimers for tubulin. The sample preparation was conducted under dark-room conditions to minimize unintended light activation, using room light filtered to block wavelengths below 580 nm (Kodak Wratten Filter No. 25). The base reaction mixture included a buffer, MgATP as an energy source, glycerol as a crowding agent, pluronic F-127 for surface passivation, and components for oxygen scavenging (pyranose oxidase, glucose, catalase, Trolox, DTT), along with ATP-recycling reagents (pyruvate kinase/lactic dehydrogenase, phosphoenolpyruvic acid). The reaction mixture consisted of 59.2 mM K-PIPES pH 6.8, 4.7 mM MgCl2, 3.2 mM potassium chloride, 2.6 mM potassium phosphate, 0.74 mM EGTA, 1.4 mM MgATP, 10\% glycerol, 0.50 mg/mL pluronic F-127, 2.9mg/mL pyranose oxidase, 3.2 mg/mL glucose, 0.086 mg/mL catalase, 5.4 mM DTT, 2.0 mM Trolox, 0.026 units/$\mu$L pyruvate kinase/lactic dehydrogenase, and 26.6 mM phosphoenolpyruvic acid.

We note that the sample is sensitive to the buffer pH and mixture incubation time. For our experimental conditions, the mixture pH is around 6.4 and we perform the experiments within 2 hours of constructing the mixture.

\begin{figure}[t]
	\centering
	\includegraphics[width=0.9\linewidth,angle=0]{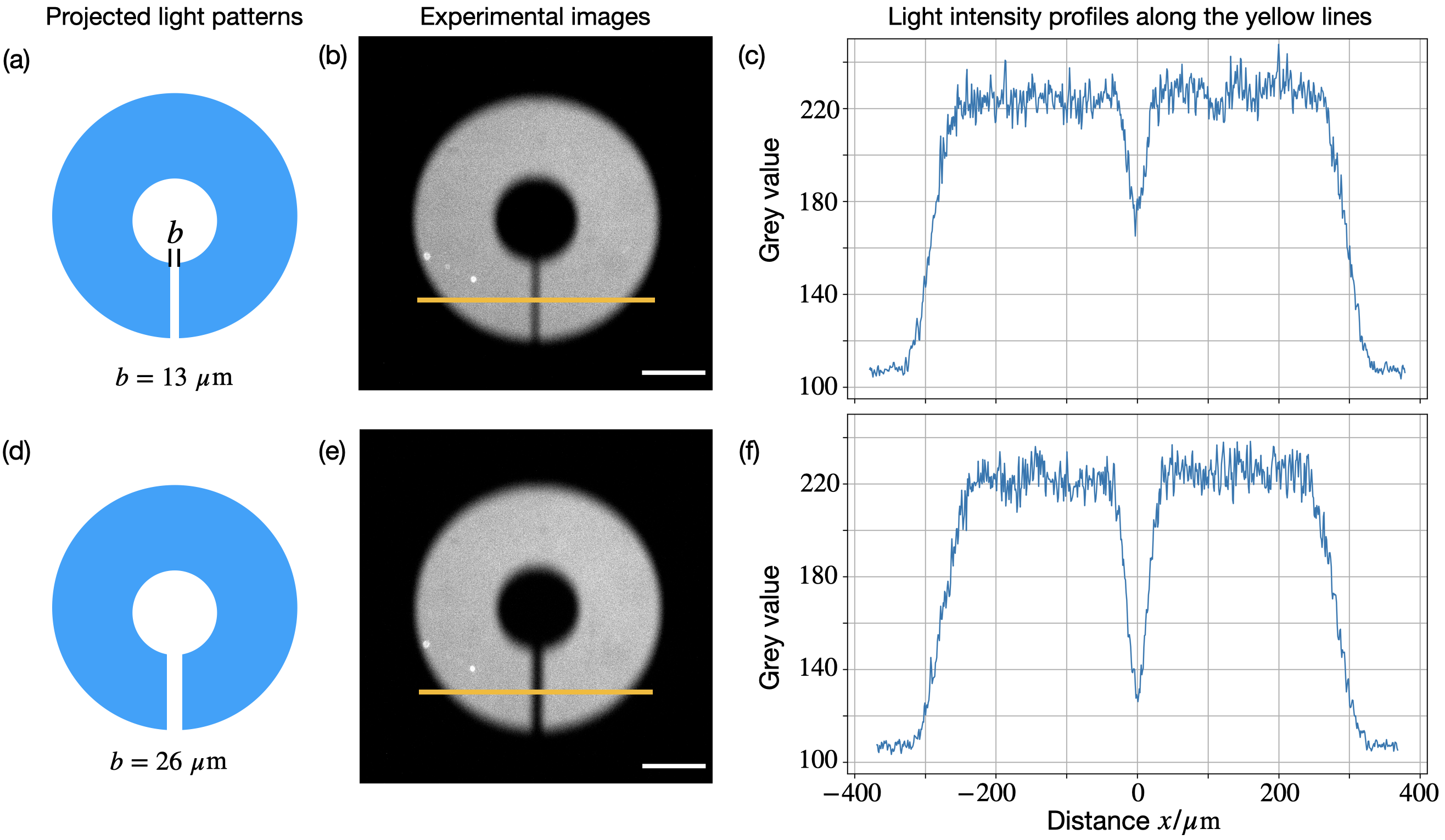}
	\caption{Light patterns for O-rings with two gap sizes. The top row (a--c) shows the 13~$\mu$m gap, and the bottom row (d--f) shows the 26~$\mu$m gap. For each case: (a, d) programmed illumination patterns used as the projector input; (b, e) corresponding microscope images of the projected patterns, with yellow lines indicating the regions used for intensity analysis; (c, f) pixel intensity profiles (gray value) along the yellow lines in (b, e). Scale bars in (b) and (e) are 200~$\mu$m.}
	\label{Fig::light}
\end{figure}

\subsection{Microscope setup}

We conducted the experiments using an automated widefield epifluorescence microscope (Nikon Ti-2), custom-modified for two additional imaging modes: epi-illuminated pattern projection and LED-gated transmitted light. Light patterns from a programmable DLP chip (EKB Technologies DLP LightCrafter™ E4500 MKII™ Fiber Couple) were projected onto the sample via a user-modified epi-illumination attachment (Nikon T-FL). The DLP chip was illuminated by a fiber-coupled 470 nm LED (ThorLabs M470L3). The epi-illumination attachment featured two light-path entry ports: one for the projected pattern light path and the other for a standard widefield epi-fluorescence light path. These light paths were combined using a dichroic mirror (Semrock BLP01-488R-25). The magnification of the epi-illumination system was calibrated to ensure the camera's imaging sensor (FliR BFLY-U3-23S6M-C) was fully illuminated when the entire DLP chip was activated. Micro-Manager software, running custom scripts, controlled the pattern projection and stage movement. For the transmitted light path, we replaced the standard white-light brightfield source (Nikon T-DH) with an electronically time-gated 660 nm LED (ThorLabs M660L4-C5) to minimize light-induced linkage during brightfield imaging.

\subsection{Light pattern construction}

In this work, we followed the illumination protocol from Ref. \cite{ross19}, using pulsed light to activate motor-linking. A light pattern is projected onto the active matter system every 10 seconds, with each pulse lasting 30 milliseconds. This protocol ensures sufficient motor activation for network contraction and sustained fluid flow. Shorter illumination periods fail to generate enough cross-linked microtubules, while longer or continuous illumination causes motor activation outside the intended region, disrupting the network and fluid flow.

For generating V-shape light patterns, we utilized a Python script to calculate the V shape's coordinates and mapped them onto a 1280 by 800 pixel canvas. The pattern was displayed as white on a black background. The final V shape pattern was saved as a TIFF file at the specified location for subsequent experimental use. A custom Java script was used to read the tiff file and project the custom light pattern to the active matter system in micro-manager.

To obtain a reference image of the projector output, the active matter mixture was replaced with a 1:100 dilution of Goat anti-Mouse IgG (H+L) Highly Cross-Adsorbed Secondary Antibody conjugated with Alexa Fluor\textsuperscript{\texttrademark} Plus 488 (Product No. A32723). The same light pattern used in the healing experiment (Fig. 1 in the main text) was projected directly onto the Alexa Fluor\textsuperscript{\texttrademark} Plus 488 solution. Imaging was performed using the identical microscope settings as those used during the healing experiments.

The light intensity profiles at the O-ring gaps are plotted in Fig. \ref{Fig::light}. At the edges of the projected light pattern, the intensity decays to the background level within a $\sim$20 $\mu$m transition layer. For a projected gap size of $b = 13$ $\mu$m, the transition regions strongly overlap, whereas for $b = 26$ $\mu$m the overlap is limited, as shown in Fig. \ref{Fig::light} (c) and (f).

\begin{figure}[t]
	\centering
	\includegraphics[width=\linewidth,angle=0]{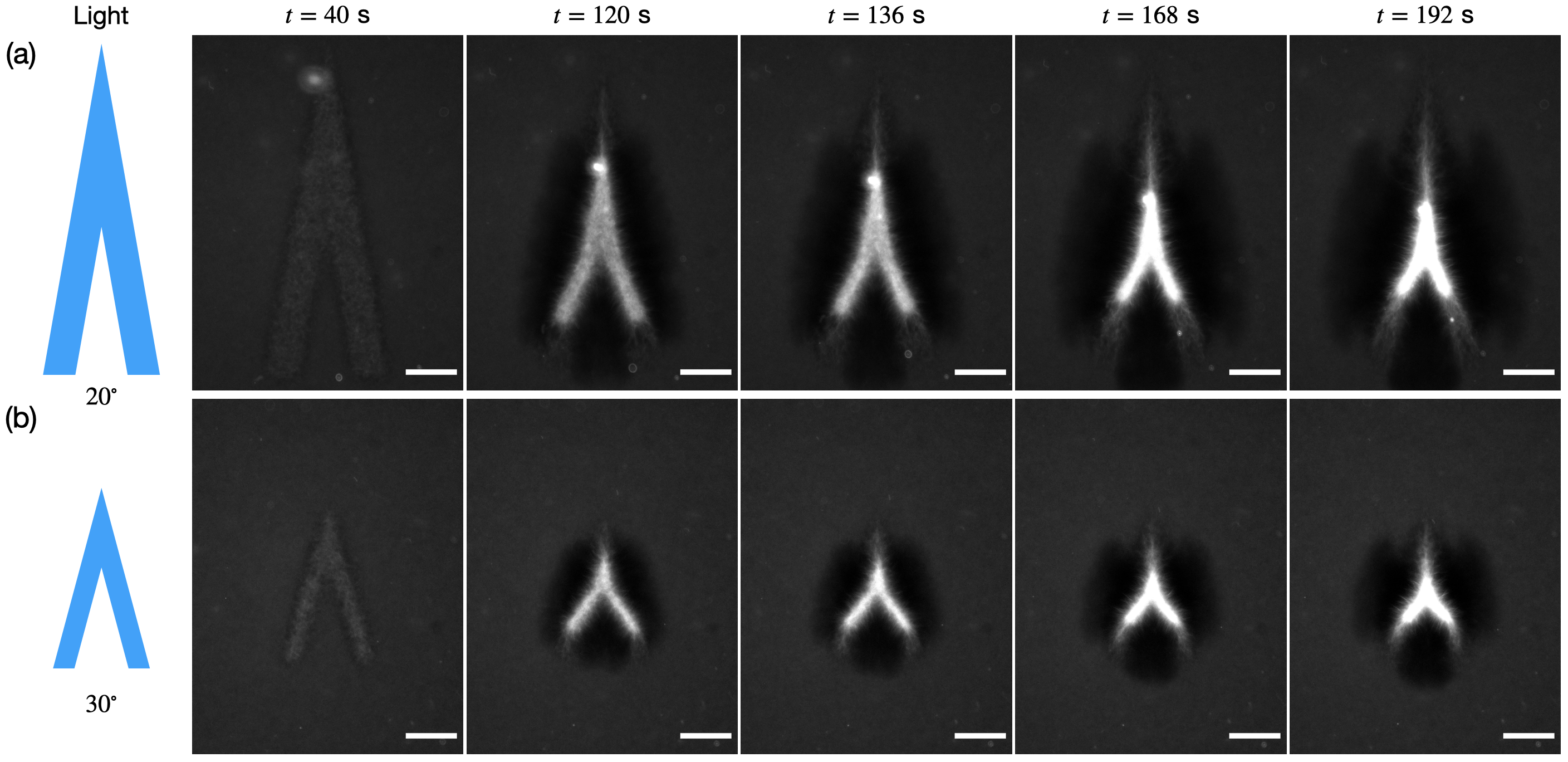}
	\caption{Experimental images of concave networks with (a) $\theta=20^\circ$ and (b) $\theta=30^\circ$. Scale bars, 100 $\mu$m.}
	\label{Fig::concave}
\end{figure}

\subsection{Image processing for centerline-curvature measurement}

We first separated the left and right arms of the networks. We used the ``skeletonize" function from the scikit-image package  in Python to extract the skeleton of each arm. We then used the FilFinder package  to pick out the longest branch of the skeleton, which is the centerline of the arm.

To measure the curvature of the centerline, we first fitted the centerline with a parabola $y= ax^2+bx+c$. The local curvature $\kappa$ of the centerline is  $\kappa = y''/(1+y'^2)^{3/2}$ and the average curvature  $[\kappa]$ is 
\begin{equation}
	[\kappa]= \frac{\int_{x_0}^{x_t} y''/(1+y'^2) \mathrm{d} x}{\int_{x_0}^{x_t}(1+y'^2)^{1/2} \mathrm{d} x},
\end{equation}
which can be numerically integrated using $y''=2a$ and $y'=2ax+b$. $x_0$ and $x_t$ are the x-coordinates of the starting and ending points of the centerline, respectively.

\subsection{Experimental images of concave networks}

We show more examples of concave V-shaped networks with small angles in Fig. \ref{Fig::concave}.

\bibliographystyle{unsrt}
\bibliography{selforg, FlowControlref, DefectMerging, ActiveHealing}

 \end{document}